\documentclass{aa}  

\usepackage{graphicx}
\usepackage{txfonts}
\usepackage{colortbl}
\usepackage[dvipsnames]{xcolor}
\usepackage{epstopdf}

\usepackage{cancel} 

\newcommand{\TDD}[1]{\textcolor{black}{#1}}

\begin{document} 

\title{Massive stars evolution with new $\rm ^ {12}C+^{12}\!C$ nuclear reaction rate -- the core carbon-burning phase}

   \authorrunning{T. Dumont et al.}
   
   \titlerunning{Massive stars evolution with new $\rm ^ {12}C+^{12}\!C$ nuclear reaction rate -- the core carbon-burning phase}
   

   \author{T. Dumont
          \inst{1} \fnmsep\thanks{e-mail: thibaut.dumont@iphc.cnrs.fr},
          E. Monpribat \inst{1},
          S. Courtin \inst{1,2},
          A. Choplin \inst{3},
          A. Bonhomme \inst{1},
          S. Ekström \inst{4},
          M. Heine \inst{1},
          D. Curien \inst{1},
          J. Nippert \inst{1},
          and G. Meynet \inst{4}
          }

   \institute{University of Strasbourg, CNRS, IPHC UMR 7178, F-67000 Strasbourg, France
         \and University of Strasbourg, Institute of Advanced Studies (USIAS), Strasbourg, France
         \and Institut d'Astronomie et d'Astrophysique, Université Libre de Bruxelles, CP 226, 1050, Brussels, Belgium
         \and Department of Astronomy, University of Geneva, Chemin Pegasi 51, 1290 Versoix, Switzerland
             }

   \date{(Received; Revised; Accepted)}

 
  \abstract
   {Nuclear reactions drive the stellar evolution and contribute to the stellar and galactic chemicals abundances. New determinations of the nuclear reaction rates for key fusion reactions of stellar evolution are now available, paving the way to improved stellar model predictions.} 
   {We explore the impact of new $\rm ^ {12}C+^{12}\!C$ reaction rates for massive stars evolution, structure, and nucleosynthesis at carbon-ignition and during core carbon-burning phase. We analyse the consequences for stars of different masses including rotation-induced mixing.
   }
   {We computed a grid of massive stars from 8 to 30 M$_\odot$ at solar metallicity using the stellar evolution code GENEC, and including the new reaction rates. We explored the results using three different references for the rates, with or without rotation. We study the effect in terms of evolution, structure\TDD{, and critical mass limit between intermediate and massive stars}. We explored the consequences for heavy-element nucleosynthesis during the core carbon-burning phase by means of a one-zone nucleosynthesis code.
   }
   {We confirm the significant impact of using the recent nuclear reaction rates following the fusion suppression hypothesis at deep sub-barrier energies (hindrance hypothesis) as well as the mass-dependent effect of a resonance at 2.14 MeV with dominant feeding of the $\alpha$ exit channel of $\rm ^ {12}C+^{12}\!C$ fusion reaction. This impacts the characteristics of the core of stars from the C-ignition and during all the core C-burning phase (temperature/density, lifetime, size, convective/radiative \TDD{core}). The change of nuclear reaction rates modifies the central nucleosynthesis of the stars during the core-carbon burning phase, resulting in an underproduction of s-process elements, especially when including the rotation-induced mixing that amplifies the effects.
   }
   {The correct and accurate determination of the nuclear reaction rates, with especially the existence and location of resonnances, impacts stellar evolution in many aspects affecting the model predictions. The choice of the nuclear reaction rates reference for the $\rm ^ {12}C+^{12}C$ fusion reaction changes significantly the behaviour of the core during the carbon-burning phase, and consequently drives changes in the nucleosynthesis and end-of-life of stars. This choice is then to be taken carefully in order to interpret stellar evolution from the super AGB phase and their massive white dwarf remnants to the core-collapse supernovae of massive stars.
   }


   \maketitle
   
%

\section{Introduction}
\label{section:introduction}

The question of the origin of the elements is an old problem in astrophysics. Since the pioneering work of \citet[][]{Burbidge1957} and \citet{Cameron1957}, some global features are now well known, but others are still debated and at the heart of active research activities. Since the Big Bang, chemical abundances are continuously evolving due to the nucleosynthesis. It involves different processes with mainly nuclear fusion or particle capture and the subsequent reactions, and it takes place in a variety of astrophysical sites, from stars to supernovae and merging events, or spallation reactions. The reconstruction of the chemical evolution of the Universe is then complex and yet-to-be fully understood. After the Big Bang, stars have been the furnace in which most of the elements heavier than lithium are synthesised. Through their winds or at the moment of their explosion, these new elements are deposited in the interstellar medium and may be served to form new stars, planets and their inhabitants. Stellar evolution codes have been developed to describe the evolution of stars and require a robust and accurate description of nuclear reactions \citep[\textit{e.g.}][]{Liccardo2018,2021FrASS...7..112V,2023A&ARv..31....1A}. In this sense, major effort has been made by the nuclear physics community to provide to stellar modellers the most accurate nuclear reaction rates. In recent years, thanks to progresses and new studies, reaction rates for fusion reactions have been measured using sophisticated approaches to reach the experimentally challenging low astrophysical energies of interest. Important progresses were made in direct and indirect methods of measurement \citep[\textit{e.g.}][]{2015PhRvL.114y1102B,2018PhRvC..97a2801J,2018Natur.557..687T,2020PhRvL.124s2702T} giving access to new nuclear reactions rates for the successive fusion burning stellar evolution phases. \\ Aside from the commonly used \citet[][hereafter CF88]{1988ADNDT..40..283C}, \citet[][Nacre]{1999NuPhA.656....3A}, and \citet[][NacreII]{2013NuPhA.918...61X}, new rates are now available or will be available in the near future \citep[\textit{e.g.}][]{2017RvMP...89c5007D,2018PrPNP..98...55B,2021PhRvL.127o2701C,2022A&A...660A..47M,2023ApJ...945...41S}.

In the present paper, we focus on the nuclear fusion reaction $\rm ^ {12}C+^{12}C$. This reaction generates a $\rm ^{24}Mg^*$ compound nucleus that eventually disintegrates into three main exit channels:
\begin{equation}
    \rm
     ^{12}C+^{12}C \xrightarrow{}^{24}Mg^* \left\{ 
     \begin{array}{lll}
                \xrightarrow{} \rm ^{20}Ne+\alpha & \text{(Q = 4.62 MeV)} \\
                \xrightarrow{} \rm ^{23}Na+p & \text{(Q = 2.24 MeV)} \\
                \xrightarrow{} \rm ^{23}Mg+n & \text{(Q = -2.62 MeV)}
    \end{array}
    \right\}
    \label{eq:channels}
\end{equation}  
For this reaction, several studies have investigated the effect of fusion hindrance \citep{2007PhRvC..76c5802G} or resonances with enhanced reaction probability and particular decay pattern, applying arbitrary factors relative to the CF88 reference \citep{2012MNRAS.420.3047B, 2013ApJ...762...31P}. 
The $\rm ^ {12}C+^{12}\!C$ reaction is indeed an important nuclear reaction taking place in different astrophysical sites: the core C-burning phase \citep[\textit{e.g.}][]{2006A&A...448..717S,2012MNRAS.420.3047B,2013ApJ...762...31P,2022A&A...659A.150D}, shell C-burning \citep[\textit{e.g.}][]{1991ApJ...371..665R,1993ApJ...419..207R,2007ApJ...655.1058T}, white dwarfs accretion and SN Ia supernovae \citep[\textit{e.g.}][]{2004ApJ...615..444S,2019ApJ...887..188A,2019MNRAS.483..263W}, explosive nucleosynthesis \citep[e.g.][]{2002RvMP...74.1015W}, and superbursts\footnote{Likely due to thermonuclear flashes occurring at the surface of neutron stars.} \citep[\textit{e.g.}][]{2002ApJ...566.1045S,2009ApJ...702..660C}. \\ The C-burning phase is, especially, a fundamental phase during massive stars evolution as the first and longest phase during which most of the energy produced in the star is released through neutrinos and no longer through radiation \citep[\textit{e.g.}][]{2002RvMP...74.1015W}. This phase impacts significantly the stellar evolution and structure, and then its end-of-life (later stages of evolution will impact as well the fate of massive stars, see for instance \citet{2016ApJ...833..124M}, \citet{2021ApJ...916...79C} and Sect.~\ref{sect:CONCLUSION}). Having accurate nuclear reaction rates is then important for computing the evolution of stars, high energy events and for unveiling the origin of the elements in the Universe. \\

The C-burning phase is the third fusion phase taking place in stellar evolution and C-ignition usually defines the so-called mass limit M$\rm _{up}$. Two other critical mass limits are important to define close to the transition between intermediate mass stars and massive stars ($\approx$ 8-10 M$_{\odot}$ at solar metallicity): M$\rm _n$ the minimum initial stellar mass for the formation of a neutron star resulting from an electron-capture supernovae, and M$\rm _{mas}$ the minimum initial stellar mass for stars that will undergo all nuclear burning phases and explode as iron core collapse supernovae. Intermediate stars initially less massive than M$\rm _{up}$ leave CO white dwarfs remnants, stars with mass between M$\rm _{up}$ and M$\rm _n$ leave either hybrid C-ONe or ONe white dwarfs remnants, stars with mass between M$\rm _{n}$ and M$\rm _{mas}$ leave a neutron star, and stars more massive than M$\rm _{mas}$ leave a neutron star or a black hole \citep[e.g.][]{Doherty2017}\footnote{As mentioned in the reference, many names are given for the different mass limits: M$\rm _{up}$=M$\rm _{CO}$, M$\rm _{n}$=M$\rm _{EC}$, and M$\rm _{mas}$=M$\rm _{up'}$=M$\rm _{up*}$}. C-ignition is typically taking place for a CO core of at least 1.05 M$_{\odot}$ at temperatures of 0.6-0.8 GK \citep[][]{1989A&A...210..155M,2007A&A...476..893S,2009pfer.book.....M}. Between M$\rm _{up}$ and M$\rm _{mas}$, the core centre is generally cooler than the close off-centre ($\le 0.1$~$\rm Mr/M_{tot}$) because of the neutrinos loss. The ignition then takes place off-centre and a flame arises before propagating toward the centre depending on the temperature. It corresponds to the so-called super asymptotic giant stars or SAGB stars that will end their life as white dwarf or in electron capture supernovae \citep[see the series of papers][and \citealt{Limongi2024ApJS}]{2006A&A...448..717S,2007A&A...476..893S,2010A&A...512A..10S}. \\ The C-burning phase is highly dependent on the second fusion phase, the He-burning phase, during which $\rm ^{12}C$ is synthesized through the triple-alpha reaction (main reaction of the He-burning phase) and then burned into $\rm ^{16}O$ at the end of the same phase through the nuclear reaction $\rm ^{12}C(\alpha,\gamma)^{16}O$. The resulting O-dominated CO core then determines the C-ignition and C-burning phase (core or off-centre), as well as the burning lifetime (from several thousand years for the less massive stars to a few hundred years for stars of 30-35 M$_{\odot}$ at solar metallicity). The main products of the C-burning phase are $\rm ^{20}$Ne and $\rm ^{24}$Mg whereas $\rm ^{23}$Na is converted into $\rm ^{20}$Ne through $\rm ^{23}Na(p,\alpha)^{20}Ne$ reaction, eventually forming a ONe core (or a hybrid C-ONe core for some SAGB stars) with presence of $\rm ^{24}$Mg and $\rm ^{23}$Na at lower abundances level. Before this phase, neutrons are produced through $\rm ^{22}Ne(\alpha,n)^{25}Mg$ at the end of the core He-burning when temperature is high enough and $\rm ^{13}C(\alpha,n)^{16}O$ (at the beginning of the He-burning phase and at the end of He-burning phase when rotation-induced mixing provides fresh $\rm ^{13}C$ from the H-burning shell), and captured for a part of them by neutron poisons elements like $^{16}$O or Fe. Reaching the C-burning phase, the resulting neutron abundance from He-burning phase will impact the nucleosynthesis including the core C-burning phase where new sources and new poisons will take place, impacting the final nucleosynthesis and efficiency of the slow neutron-capture process or s-process \citep[\textit{e.g.}][]{2008ApJ...687L..95P,2010ApJ...710.1557P,2013ApJ...762...31P,2016MNRAS.456.1803F,2018A&A...618A.133C}. \\

In a previous paper, \citet[][hereafter M22]{2022A&A...660A..47M} tested these new nuclear reaction rates for two classical non-rotating models of 12 and 25 M$_{\odot}$ at solar metallicity. In the present work, we aim at extending this work exploring a larger range of stellar masses at solar metallicity, and exploring as well the effects of rotation.\\
In Sect.~\ref{GENEC} we describe the input physics of the Geneva stellar evolution code (GENEC) used for this work. We present the context in which the new nuclear reaction rates determination by experimental studies for $\rm^{12}C+^{12}C$ takes place in Sect.~\ref{sect:nucl}. In Sect.~\ref{sect:class}, predictions from classical and rotating models at different masses are described. In Sect.~\ref{sect:massdep} we explore the impact on mass limits and discuss the impact on the end-of-life of stars resulting from stellar structure changes. Finally, we describe the results for nucleosynthesis in Sect.~\ref{sect:nucleo}, and in Sect.~\ref{sect:CONCLUSION}, we summarise our results, discuss model predictions, and conclude. 

\section{The stellar evolution code}
\label{GENEC}
In this work, we used the stellar evolution code GENEC as described in \citet[][]{2008Ap&SS.316...43E,2012A&A...537A.146E}, which we refer to for details and references. We computed a grid of models at solar metallicity Z = 0.014 ([Fe/H] = 0), with initial masses of 8, 9, 10, 12, 15, 17, 20, 22, 25, 26, and 30 M$_{\odot}$, with or without rotation. When including rotation we consider an initial ratio of surface velocity to critical velocity of $\rm v_{ini}/v_{crit}$=0.4 that corresponds to the peak of velocity distribution of a sample of B stars observed by \citet{2010ApJ...722..605H}. Our models are computed from the zero age main-sequence (ZAMS) to the Ne-ignition, defined where the neon core abundance $X_{^{20}Ne}$ is decreased by $3\times10^{-3}$ (in mass fraction) compared to the $X_{^{20}Ne}$ value at C-exhaustion. When \TDD{models do not reach the temperature for Ne-ignition of $\approx$ 1.2-1.5 GK \citep{2009pfer.book.....M}, they are computed until the onset of the thermally pulsating phase after the C-exhaustion (defined when $X_{^{12}C}\le 10^{-5}$), or when reaching the carbon flash}.

\subsection{Input physics}
\label{sub:inputphysics}
The models presented in this work were computed with the same inputs physics (equation of state, opacities, nuclear reactions rates, grey atmosphere) as in M22 except specified otherwise.
For all models, convection is treated using the mixing length theory with a parameter $\alpha_{\rm MLT}=1.6$ adjusted from model calibrations on the Sun with abundances\footnote{We use initial abundances for H and He as X = 0.720 and Y = 0.266 from the same calibration.} from \citet{2005ASPC..336...25A} and Ne enhancement following \citet{2006ApJ...647L.143C}, in agreement with the recent works by \citet{2018ApJ...855...15Y,2021A&A...653A.141A}. We assume the Schwarzschild criteria for convective stability and apply a core overshoot for stars exhibiting a convective core along evolution (masses equal or above 1.2 M$_{\odot}$ at solar metallicity), and until the end of the He-burning phase; with an overshoot parameter $\rm d_{\rm over}=0.1\,H_p$, and where $\rm H_p$ is the pressure scale height.

\subsection{Nuclear reaction rates: the complex pattern of $\rm ^ {12}C+^{12}\!C$}
\label{sect:nucl}

In the actual thermal conditions of C-burning in massive stars, the energies of the carbon nuclei in the $\rm ^{12}C+^{12}\!C$ fusion reaction are far below the Coulomb barrier from the repelling charges. At energies around the barrier, low lying states in $\rm ^{12}C$ contribute to the reaction with individual $Q$-value and cross section \citep[]{back2014PRM}. At deep sub-barrier energy, suppressed fusion was observed for many fusing nuclei \citep[]{jiang2021EPJA} and the effect in the $\rm ^{12}C+^{12}\!C$ system is still under debate as precise measurements of cross sections (below nanobarn) need to be carried out. In addition, resonance structures (strongly enhanced cross sections) are apparent over the entire energy range explored so far. Those might indicate clustering effects in the compound system \citep[]{bohr1936nat, ikeda1968} with possible impact on the fusion dynamics \citep{vogt1960, taras1978, adsley2022}.

Rates for the $\rm ^{12}C+^{12}\!C$ fusion reaction have been widely studied in recent years due to its complex nuclear behaviour and strong oscillations in the cross section often attributed to $\rm ^{12}C-^{12}C$ cluster resonances in $\rm ^{24}Mg$ \citep[][]{Jenkins_2015}. Until now two main features have been identified. First: a general unexpected steep-fall of the cross section at low energies attributed to the fusion-hindrance phenomenon \citep[][]{2004PhRvL..93a2701J,Jiang2007}; second: the observation of one or several resonances at different energies \citep[][Nippert et al. 2024 in prep]{2007PhRvL..98l2501S}. Using the so-called Trojan horse method formalised by \citet[][]{2008JPhG...35a4016M}, \citet{2018Natur.557..687T} achieved the indirect determination of the reaction rates from 2.7 MeV to 0.8 MeV and showed a strong increase of the rates (several orders of magnitude compared to CF88 rates) at low energies, highlighting the complexity characterising this reaction. However, several uncertainties have been highlighted by \citet[][]{2019PhRvC..99f4618M} for the use of this indirect method in the case of carbon as well as the strong sensitivity of the results to parameters and hypothesis of the method, and these results are still debated. In parallel, direct determinations of the rates have been achieved down to low energies exhibiting the complex pattern of the $\rm ^ {12}C+^{12}\!C$ fusion reaction by \citet{2015PhRvL.114y1102B,2018PhRvC..97a2801J,2020PhRvL.124s2702T}. In particular, thanks to a new direct measurement using a particle-gamma coincidence technique, the STELLA team was able to probe down to a low energy in the centre of mass of $\rm E_{cm} \approx 2$~MeV \citep{2020PhRvL.124s2701F}, and determined new reliable and accurate reaction rates for $\rm ^{12}C+^{12}\!C$ (M22).\\ 
To append to the previous paragraph, a synthetic review on the very recent experimental progresses on the $\rm ^{12}C+^{12}\!C$ system using indirect and direct techniques can be found in \textit{e.g.} \citet[][]{Courtin_2023}.
\newline
 \\ In GENEC, we use the numerical tables from the NACRE compilation \citep{1999NuPhA.656....3A} generated with the NextGen web interface. If not available in NACRE, rates are taken from CF88 or from nuclear reaction dedicated studies as described in Sect.~2.2 by \citet{2008Ap&SS.316...43E} and Sect.~2.2 by \citet{2012A&A...537A.146E}. \TDD{GENEC takes into account a network of nuclear reactions involving 32 stable and unstable species from $^1$H to $^{56}$Ni plus neutrons, see details in Tab.~\ref{tab:annexe0}.} It allows tracking the reactions that contribute significantly to nuclear energy \TDD{during all evolution} and the nucleosynthesis of the main elements to be followed \TDD{in H-burning and He-burning phases}. In this work we updated the reaction rates for the $\rm ^ {12}C+^{12}\!C$ nuclear reaction according to M22 for the $\alpha$ and proton channels, and according to \citet{2015PhRvL.114y1102B} for the neutron channel (see Eq.~\ref{eq:channels}). In the present version of GENEC, the channels are summed as a single $\alpha$ channel as described in M22. This approximation is valid to explore the structure and general evolution of stars but it is not any more the case when focusing on nucleosynthesis (since only $\rm ^{20}Ne$ is produced)\footnote{An updated version of GENEC will be used in a future work to account for the different channels self-consistently along evolution. However, the computation time will be increased significantly and cannot be currently used for a full grid of models as presented here.}. In addition to the multi-zone stellar evolution code GENEC, we then use a one-zone nucleosynthesis code including a larger network of nuclear reactions and 736 species from $^1$H to $^{212}$Po \citep[\textit{e.g.}][]{2016A&A...593A..36C, Frost-Schenk22} in order to study the potential effects of the different channels on the production of heavy elements. We note that the one-zone code does not take into account the effects of mixing. However, \citet{2016A&A...593A..36C,Frost-Schenk22} showed that injecting $^{13}$C and $^{14}$N at a rate of $2.5\times 10^{-7}$ M$_{\odot}$.yr$^{-1}$ (as calibrated thanks to rotating stellar models computed with GENEC) to mimic the transport of these chemical species by rotational mixing was able to reproduce very well the effect of rotation-induced mixing during the He-burning phase (see their Fig.~5).  
 In the present case, due to the short duration of the core C-burning phase (see Tab.~\ref{tab:lifetime}), we assume that rotational mixing barely operate during this stage and therefore do not inject any chemical species as mentioned above.  
 Nevertheless, we plan to further investigate this assumption in a future work, especially regarding the convective mixing \TDD{for which the convective turnover timescale might be shorter than the nuclear burning timescale} and the case of the lowest mass stars for which the core C-burning lifetime reaches a few thousand years. We eventually computed GENEC models with the \citet[][CF88]{1988ADNDT..40..283C} rates, within the fusion-hindrance hypothesis (HIN), and within the Hindrance+Resonance hypothesis (HINRES), see M22 for details.

\subsection{Angular momentum evolution and rotation-induced mixing}
\label{sub:stellarrotation}
Stellar rotation is implemented in GENEC as described by \citet{2008Ap&SS.316...43E}, and the shellular rotation hypothesis developed by \citet{1992A&A...265..115Z}, \citet{1998A&A...334.1000M}, and \citet{2004A&A...425..229M} is followed to describe the differential rotation in stars. The transport of angular momentum and chemicals is driven by the combined action of meridional circulation (as an advective process for angular momentum), and turbulent shear (vertical and horizontal). Diffusion coefficients are taken from \citet{1992A&A...265..115Z} and \citet{1997A&A...321..134M} for the horizontal diffusivity $\rm D_h$ and the vertical diffusivity $\rm D_v$, respectively. \\
The chemical evolution for each element $i$ is described by the general diffusion equation as:
\begin{equation}
    \rho \frac{\partial X_i}{\partial t} = \frac{1}{r^2}\frac{\partial}{\partial r} \left(r^2 \rho D \frac{\partial X_i}{\partial r}\right)+ m_i \left[ \sum_j \lambda_{ji} - \sum_k \lambda_{ik} \right],
    \label{eqdiffu}
\end{equation}
where $\rho$ is the density; $X_i$ refers to the mass fraction of element $i$; $r$ is the radius; $D = \sum_j D_j$ is the total coefficient for turbulent diffusion, written as the sum of the $j$ different diffusion coefficients describing turbulent processes such as shear, or any other process; $m_i$ is the mass of nuclei $i$; and $\lambda_{ij}$ the reaction rate producing nuclei $j$ from nuclei $i$.\\
The transport of angular momentum obeys the advection-diffusion equation:
\begin{equation}
    \rho \frac{d}{dt}(r^2 \Omega) = \frac{1}{5 r^2} \frac{\partial}{\partial r} (\rho r^4 \Omega U_2) + \frac{1}{r^2} \frac{\partial}{\partial r}\left(\nu_v r^4 \frac{\partial \Omega}{\partial r}\right),
    \label{eq:rot}
\end{equation}
where $\rho$, $r$, $\Omega$, $U_2$, and $\nu_v$ are the density, the radius, the angular velocity, the radial component of the meridional circulation velocity, and the viscosity of the vertical shear turbulence, respectively. The quantities describing the rotation ($U_2$, $\rm \nu_v$, $\rm \nu_h$) are computed at each time-step according to the formalism of \citet{1998A&A...334.1000M} \citep[for more details, see e.g. Sect.~2 by][]{Decressin2009}. The approach in GENEC is different from the approaches using free parameters like \textit{e.g.} $f_{\mu}$ and $f_c$ \citep[see][for a detailed discussion]{Nandal2023}. The explicit account in GENEC of the strong horizontal diffusion ($\rm \nu_h$) and the treatment of the transport of the angular momentum by the meridional currents through an advective equation avoids the introduction of these two free parameters. On the other hand, the precise value of $\rm \nu_h$ is still uncertain and various expressions have been proposed \citep{1992A&A...265..115Z,2003A&A...399..263M,2004A&A...425..243M,2018A&A...620A..22M}. Here we chose the one suggested by \citet{1992A&A...265..115Z}.

\subsection{Mass loss}
During their evolution, massive stars witness an important mass loss through stellar winds. In addition, as the star is rotating, an important amount of angular momentum is lost along evolution of massive stars. It is a key aspect of massive stars evolution and several prescriptions were proposed in the literature.\\ In the present study, we follow \citet{2012A&A...537A.146E}. On the main-sequence (MS), mass loss is defined following the formula given by \citet{2001A&A...369..574V}, and by \citet{1988A&AS...72..259D} in the not covered domains. \\ Along the giant phases, we follow the formula by \citet{1975MSRSL...8..369R,1977A&A....61..217R} with $\eta=0.5$ for stars of M$\rm \le12 M_{\odot}$ and the prescription from \citet{1988A&AS...72..259D} for stars of $\rm M\ge15M_{\odot}$.

\section{Stellar evolution: Core carbon-burning phase}
\label{sect:class}
\begin{figure}[t]
    \center
    \includegraphics [trim = 15mm 10mm 5mm 5mm,clip,width=95mm]{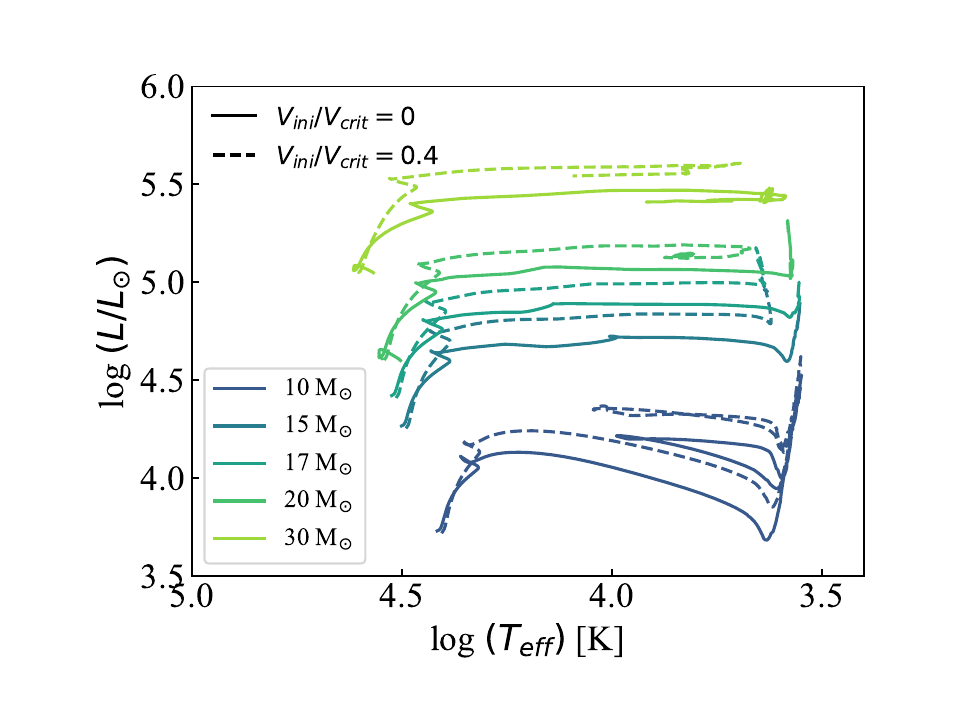}
    \caption{Evolutionary tracks in the Hertzsprung-Russell diagram for classical models (full lines) and rotating models (dashed lines) of 10, 15, 17, 20, and 30 M$_{\odot}$.}
    \label{fig:HRD}
 \end{figure}

In this section, we present results for classical and rotating models, exploring the impact of the new rates for the general stellar evolution and carbon-ignition/carbon-burning (hereafter C-ignition/C-burning) characteristics, especially focusing on the dependence to the initial mass. \\
Figure~\ref{fig:HRD} shows the evolutionary tracks in the Hertzsprung-Russell diagram for classical and rotating models of the 10, 15, 17, 20, and 30 M$_{\odot}$, at solar metallicity. As already studied by \textit{e.g.} \citet[][]{charbonnel_zahn_siess_2008,2009pfer.book.....M,2013ApJ...772..150J,2013ApJ...764...21C,2021FrASS...8...53E}, varying the mass and/or considering rotation lead to very different physical characteristics for stars from the colder intermediate mass stars ($\approx$ 8 M$_{\odot}$) to the more luminous and hotter massive stars ($\approx$ 30 M$_{\odot}$) of our grid.
\subsection{Classical models}
\label{sect:class_ana}
\begin{figure}[t]
         \center
         \includegraphics [width=88mm]{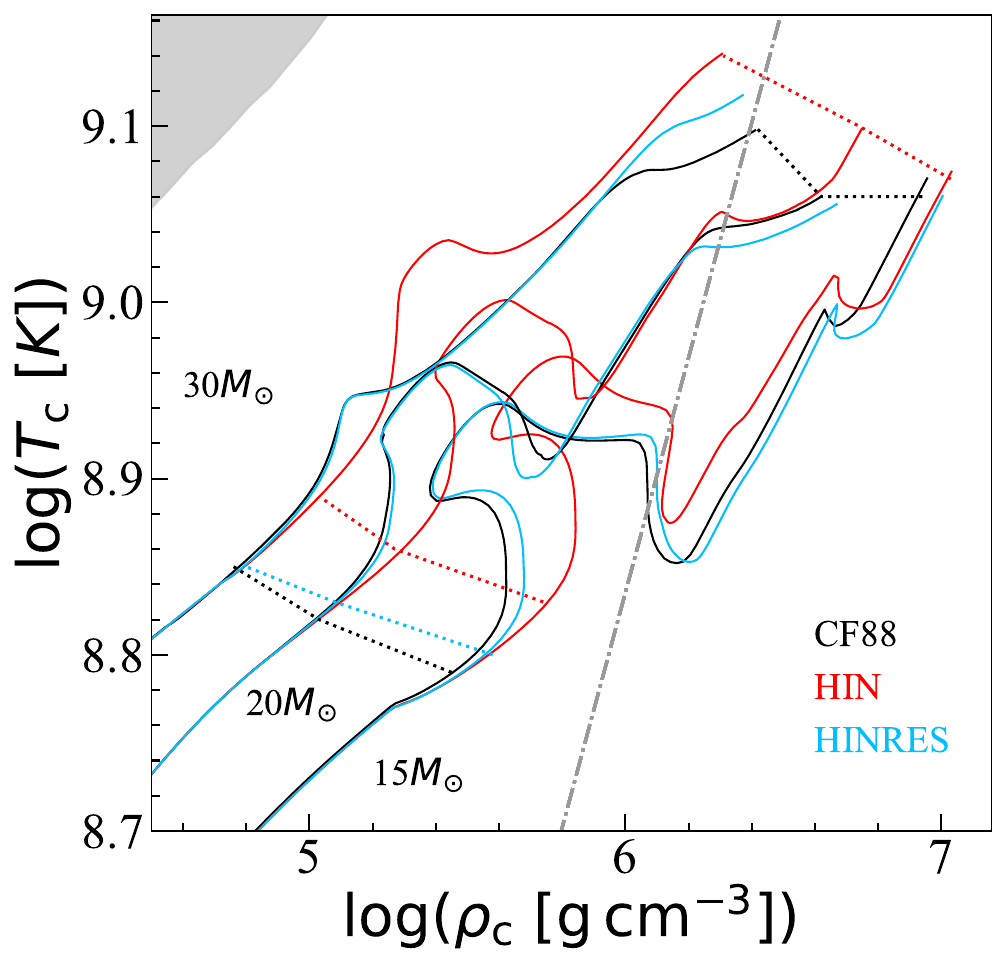}
         \caption{Evolution of T$_{\rm C}$ and $\rm \rho_{C}$ for classical models of the 15, 20, and 30 M$_{\odot}$ at solar metallicity, and for the three nuclear rates references (colour-coded). The coloured dotted lines indicate the C-ignition and core C-exhaustion for each model. The grey dotted-dashed line indicates the limit between ideal gas (left) and degenerate gas (right). The grey shaded area indicates the pair instability domain $\rm e^+e^-$.}
         \label{fig:rhot}
 \end{figure}

\begin{figure}[t]
         \center
         \includegraphics [trim = 14mm 10mm 5mm 5mm,clip,width=95mm]{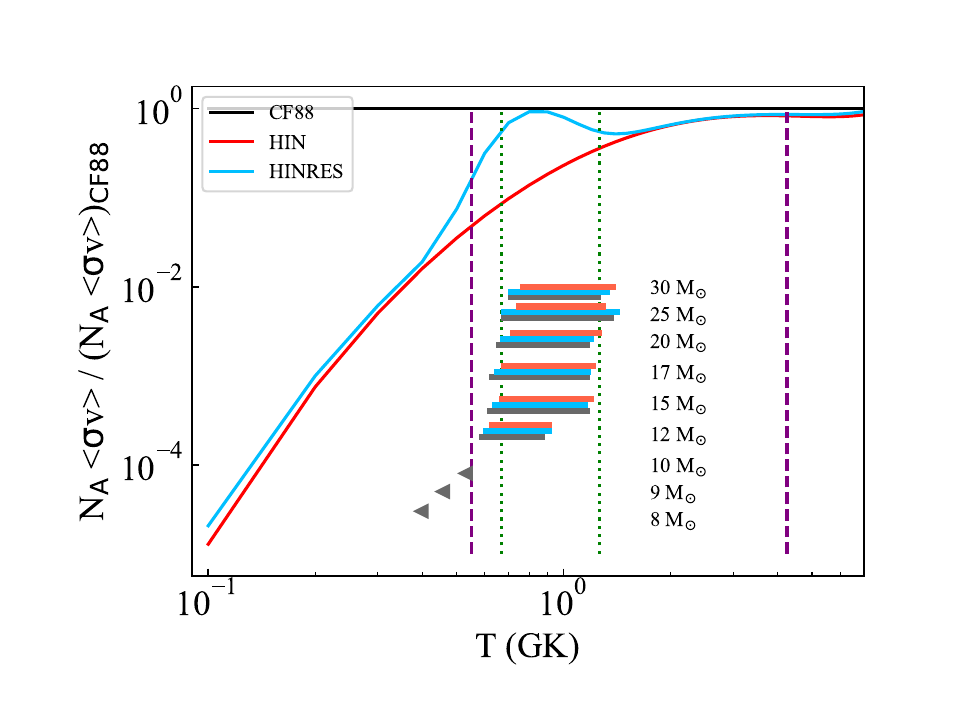}
         \caption{Reactions rates following CF88, HIN, and HINRES reference in black, red, and blue, respectively, and normalised to CF88 rates ($N_A <\sigma v>_{CF88}$). Black, red, and blue rectangles show the temperature regions where carbon fusion takes place for classical (non-rotating) models of 12, 15, 17, 20, 25, and 30 M$_{\odot}$. Black triangles give the maximum \TDD{central} temperature reached by the 8, 9, and 10 M$_{\odot}$ stars. 
         Green dotted vertical lines indicate where the HINRES rates are decreased by 45$\%$ compared to the CF88 rates. Purple dashed lines gives the area explored by the STELLA experiment.}
         \label{fig:crosstemp}
 \end{figure}

We first focus on classical non-rotating models, \textit{i.e.} models that include only convective transport as internal transport process (a core overshoot is also applied in all models). We explore the direct effect of changing the nuclear reaction rates reference on the core, and on the C-burning lifetime. Figure \ref{fig:rhot} shows the \TDD{central} temperature (T$_{\rm C}$) as a function of the \TDD{central} density ($\rm \rho_{C}$) for stars of 15, 20, and 30 M$_{\odot}$ (see also stars of 12, 17, and 22 M$_{\odot}$ in Fig.~\ref{fig:rhot_bis}). The models of our grid predict that stars $\leq$ 8 M$_{\odot}$ do not ignite carbon, stars > 10 M$_{\odot}$ ignite carbon in their centre, and stars between 9 and 10 M$_{\odot}$ ignite carbon off-centre (see \textit{e.g.} the classical 10 M$_{\odot}$ in Fig.~\ref{fig:Tcrhoc10}). They are consequently not shown here and will be discussed in Sect.~\ref{sect:offcentre}. For stars with an initial mass of 12 M$_{\odot}$ or more, as observed in M22, the HIN models (and at a lower level the HINRES models) predict higher temperatures and densities in the burning core than for CF88. It is due to the equilibrium re-adjustment of the star, in consequence of the lower cross sections  \citep[as also predicted in \textit{e.g.}][]{2012MNRAS.420.3047B,2013ApJ...762...31P}. This trend is observed for each mass with a clear shift from HIN models compared to CF88 and HINRES models at each step of C-burning phase. \\ Concerning the CF88 and HINRES models, differences are much smaller, and the change of rates affects differently the stars depending on their mass, as observed in Figs.~\ref{fig:rhot} and \ref{fig:rhot_bis}, and illustrating by C-ignition and C-exhaustion dotted lines. Firstly, the difference is predicted at the beginning of the core C-burning phase for stars between 12 and 20 M$_{\odot}$\footnote{For instance, at C-ignition: the 15 M$_{\odot}$ model predicts an increase of +0.01 dex ($\approx 0.014$ GK) and +0.13 dex in log($\rm T_c$) and log($\rm \rho_c$), respectively; the 17 M$_{\odot}$ model predicts an increase of +0.015 dex ($\approx 0.022$ GK) and +0.06 dex in log($\rm T_c$) and log($\rm \rho_c$), respectively.}, and then it is again predicted at the end of the C-burning phase especially for stars more massive than about 20 M$_{\odot}$ (with a maximum difference observed for the 25 and 30 M$_{\odot}$ models \footnote{The 30 M$_{\odot}$ model (not shown for sake of clarity) predicts an increase of +0.22 dex ($\approx 0.065$ GK) and -0.04 dex in log($\rm T_c$) and log($\rm \rho_c$), respectively, at C-exhaustion ($\rm X_{^{12}C} < 10^{-5}$)}. \\ These differences can be interpreted thanks to Fig.~\ref{fig:crosstemp} that gives the reactions rates following CF88, HIN, and HINRES references normalised on CF88 rates ($\rm N_A<\sigma v>_{CF88}$). Figure~\ref{fig:crosstemp} gives also the \TDD{central} temperatures reached by each model during the C-burning phase (or maximum \TDD{central} temperature for stars $\le$ 10 M$_{\odot}$). Especially, it shows that the effect of the resonance at about 0.8-0.9 GK (log ($\rm T_c$) = 8.90-8.95) is close to the temperature regime reached by the stars in our grid during their core C-burning phase. 
In the same figure, to guide the eye, we indicate by the green dotted vertical lines the temperature range where there is less than 45$\%$ of difference between CF88 and HINRES rates. We assume that in this range, the difference is too small to impact significantly stellar evolution/structure. This region is similar to the C-burning temperature of the 20~M$_{\odot}$ stars. It explains why this star is only slightly impacted by the change of nuclear reference from CF88 to HINRES rates that are close or identical in this region. As observed in Fig.~\ref{fig:rhot}, it shows that the main effect of the resonance, when compared to CF88, takes place outside the green dotted vertical lines zone for stars reaching colder or hotter temperatures. In other words, stars less massive than the 20 M$_{\odot}$ are impacted at the beginning of the C-burning phase (cold limit) and stars more massive than 20 M$_{\odot}$ are impacted at the end of the C-burning phase (hot limit). \\
Finally, table~\ref{tab:lifetime} gives the core C-burning lifetime\footnote{The core C-burning lifetime is defined between the C-ignition (defined when $\rm X_{^{12}C}$ decreased by 0.003 in the centre after He-exhaustion) and the C-exhaustion (defined when $\rm X_{^{12}C}<10^{-5}$).} as a function of the initial mass for the different nuclear rates references, for classical (and rotating models). As observed in M22, the HIN models have a reduced C-burning lifetime compared to the CF88 models ($\approx - 65\%$), and HINRES models ($\approx - 15\%$ compared to CF88). Indeed, reducing the nuclear reaction rates drives a decrease of the produced energy in the burning core. The star then re-adjusts itself to reach a new equilibrium by the contraction of the stellar core. This contraction leads to an increases of the \TDD{central} temperature resulting in shorter C-burning lifetimes in HIN/HINRES models. In addition, as already mentioned in M22, the lower nuclear rates of the HIN and HINRES models lead to a delayed C-ignition at higher density. It is easily observed in Fig.~\ref{fig:rhot} (dotted lines), where we can see as well that this shift is stronger between CF88 and HINRES for the less massive stars due to the location of the resonance, as already mentioned above. \\
 \begin{table*}[t!]
    \centering
    \caption{Core C-burning lifetime (in yrs) for the different masses and nuclear reaction rates.}
    \begin{tabular}{c|c|c|c|c|c|c|c|c|c|c|c|c|c}
         Mass M$_{\odot}$ & 12 & 15 & 17 & 20 & 22 & 24 & 25 & 26 & 27 & 28 & 29 & 30 & 32 \\
         \hline \hline
         Model & \multicolumn{12}{c}{Classical - Z$_{\odot}$} \\
         \hline
         CF88 & 11937 & 5106 & 3577 & 1309$^{\circ}$ & 764$^{\dagger}$ & - & 518$^{\circ}$ & 406 & 372$^{\star}$ & 331 & 281 & 242 & - \\
         HIN & 5293 & 2052 & 1362 & 470 & 254 & - & 159 & 119 & 109 & 106 & 70 & 61 & 51 $^{\star}$ \\
         HINRES & 10899 & 4579 & 3210 & 1120 & 605 & - & 384 & 352 & 303 & 273 & 244$^{\star}$ & 209 & - \\
         \hline \hline
         Model & \multicolumn{12}{c}{Rotation - Z$_{\odot}$} \\
         \hline
         CF88 & 6106 & 3067 & 1938 & 940$^{\dagger}$ & 676 & 589 & 283 $^{\star}$ & 267  & -  & -  & -  & 155 & - \\
         HIN & 2602 & 1190 & 591 & 296 & 216 & 154 & 101 & 84 & 67 & 59 $^{\star}$ & -  & 34 & -   \\
         HINRES & 5634 & 2452 & 1679 & 890 & 521 & 330 & 230$^{\star}$ & 140 & -  & -  & -  & 100 & - \\
         \hline
    \end{tabular}
    \label{tab:lifetime}
    \tablefoot{Mass limit for models showing a C-burning convective core (for masses smaller than this limit) and a radiative one (equal or above this limit). $\star$ This work. $\dagger$ According to \citet{2004A&A...425..649H} using the CF88. $\circ$ According to \citet{2021ApJ...916...79C} using the CF88 ($\approx$ 25 M$_{\odot}$) and THM ($\approx$ 20-21 M$_{\odot}$) reference rates, respectively.}
\end{table*}
\newline
This first part on the analysis about classical models confirms the general agreement on the effect of changing the nuclear reaction rates of $\rm ^{12}C+^{12}\!C$ for the core C-burning phase observed by \textit{e.g.} \citet{2007PhRvC..76c5802G,2012MNRAS.420.3047B,2013ApJ...762...31P}. Lower nuclear reactions rates, like with the HIN model, lead to an increase of the temperature and density in the core in order to re-adjust the equilibrium of the star. The C-ignition is then shifted in time and at higher temperature and density, and the C-burning phase lasts for a shorter period of time compared to the CF88/HINRES models. In addition, differences are also observed between the CF88 and HINRES models affecting as well the core conditions of stars and the behaving of C-burning phase. In particular, the temperature location of the resonance ($\approx$ C-burning temperatures of the 20 M$_{\odot}$) is significant as it drives a mass dependence observed between the lowest mass stars, impacted at the beginning of the C-burning phase, and the most massive stars, impacted at the end of the C-burning phase.

\subsection{Evolution of rotating models}
\label{sect:rot}

\begin{figure}[t]
         \center
         \includegraphics [width=90mm]{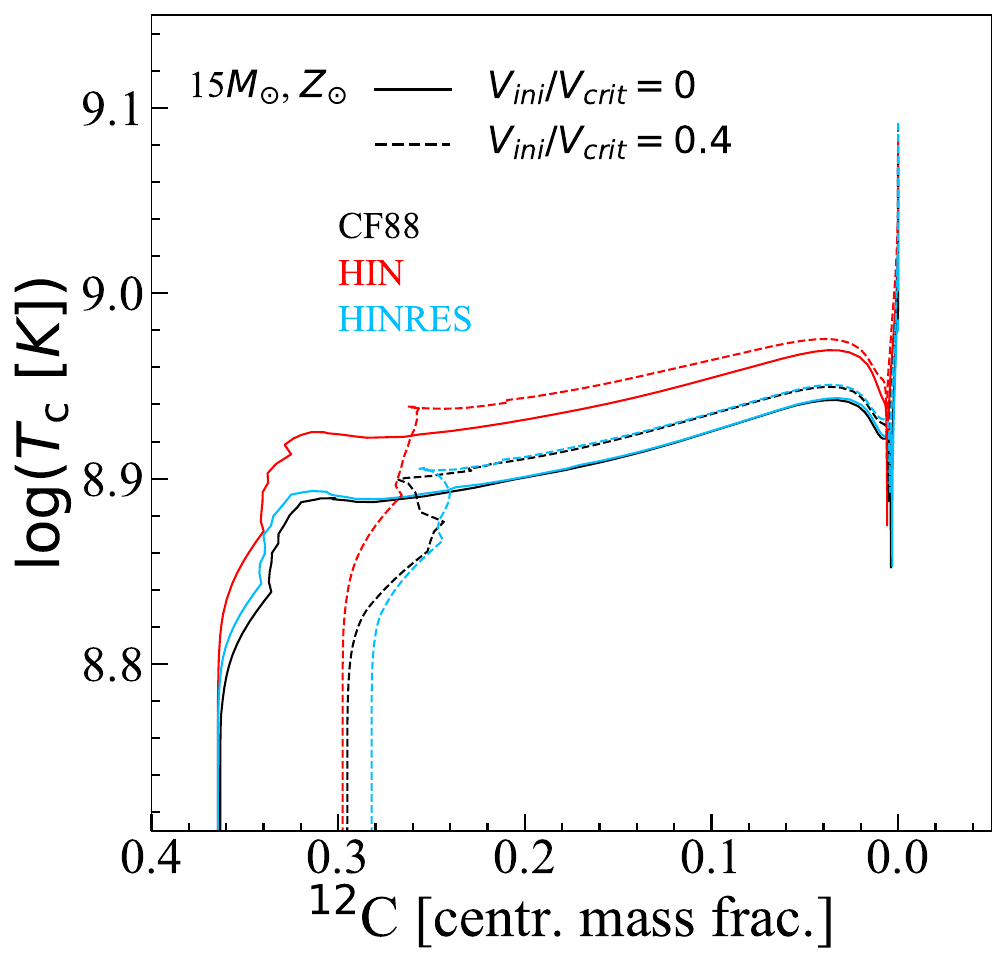}
         \caption{\TDD{Central} temperature as a function of core $\rm ^{12}C$ abundance for classical and rotating models of a 15 M$_{\odot}$ with different nuclear rates references (colour-coded).}
         \label{fig:Tc_C12rot}
 \end{figure}

\begin{figure}[t]
         \center
         \includegraphics [width=90mm]{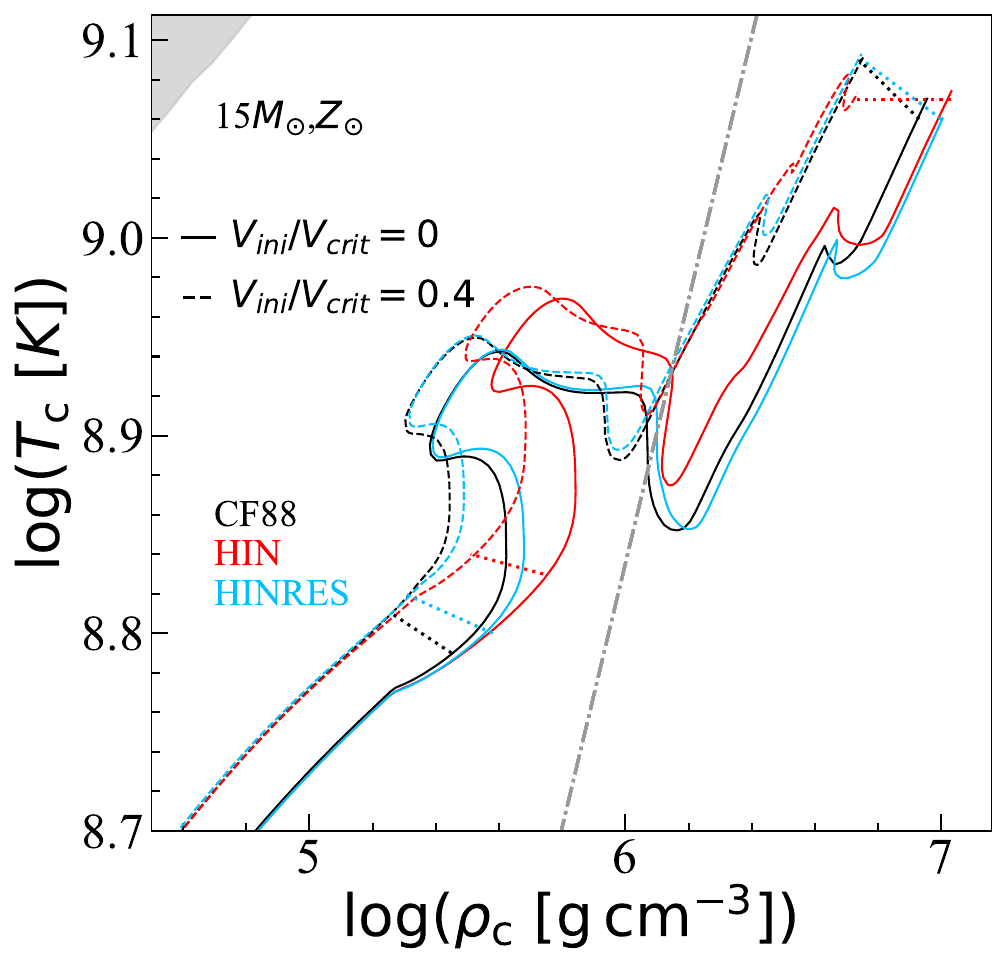}
         \caption{Same as Fig~\ref{fig:rhot} for classical and rotating models of a 15 M$_{\odot}$ star.}
         \label{fig:Tcrhoc15}
 \end{figure}

\begin{figure}[t]
         \center
         \includegraphics [trim = 14mm 10mm 5mm 5mm,clip,width=95mm]{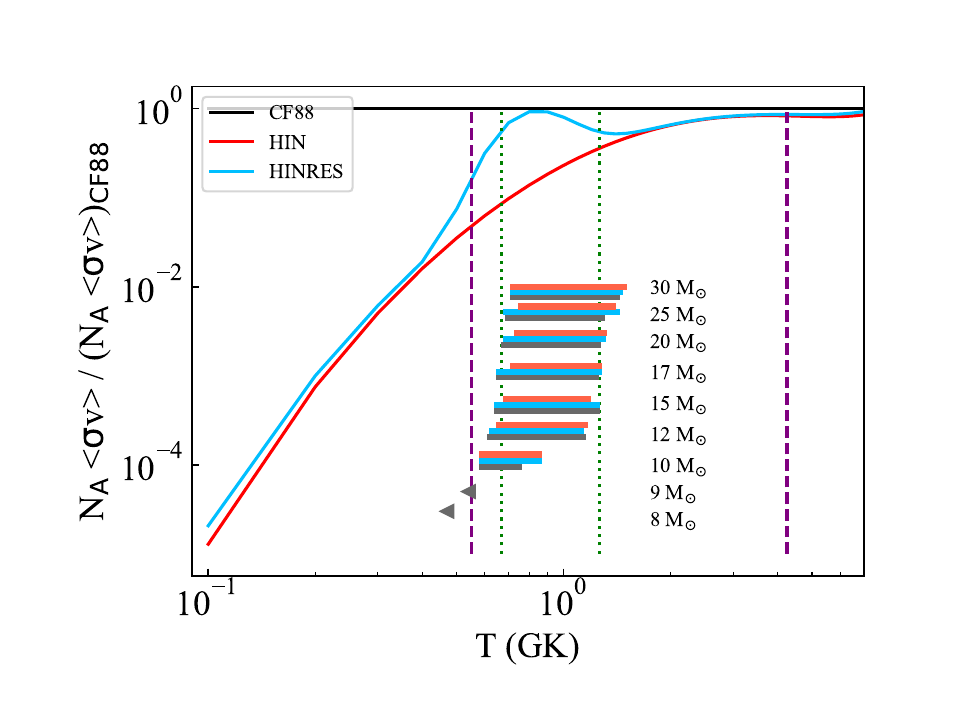}
         \caption{Same as Fig.~\ref{fig:crosstemp} for rotating models. The 10 M$_{\odot}$ now ignites carbon in core.}
         \label{fig:crosstemp_rot}
 \end{figure}

Models accounting for rotation are well known to drive important changes in model predictions due to an additional transport of both the angular momentum and the chemicals, impacting the evolution and structure of stars \citep[\textit{e.g.}][]{2021FrASS...8...53E}. The models of our grid predict that stars < 8 M$_{\odot}$ do not ignite carbon, stars > 9 M$_{\odot}$ ignite carbon in their centre, and stars between 8 and 9 M$_{\odot}$ ignite carbon off-centre. Rotation drives in each model a slightly hotter and more dense core along evolution (see Figs.~\ref{fig:Tc_C12rot} and \ref{fig:Tcrhoc15}), resulting from an increase of the core size during the MS. C-ignition is shifted in time and the ratio of $^{12}$C over $^{16}$O when reaching the ignition is smaller\footnote{We note that this ratio also depends on the initial mass with a smaller ratio for more massive stars with larger \TDD{central} temperatures.} due to a stronger efficiency of the reaction $^{12}$C($\alpha$,$\gamma$)$^{16}$O at the end of a hotter He-burning phase. The C-burning lifetime is then significantly shortened compared to classical models (see Tab.\ref{tab:lifetime}) due to a smaller amount of fuel.  
\newline
Figure~\ref{fig:Tcrhoc15} shows the \TDD{central} temperature as a function of \TDD{central} density with/without rotation for the 15 M$_{\odot}$. From the analysis achieved for the classical models in Sect.~\ref{sect:class_ana}, the general trends observed when changing the nuclear reaction rates reference are similar with rotation. The lower cross sections lead to higher temperatures and densities in the \TDD{centre} (see Fig.~\ref{fig:Tc_C12rot}) and to an even shorter lifetime of the C-burning phase as presented in Tab.~\ref{tab:lifetime} ($\approx -68\%$ and $\approx -18\%$ for the HIN and HINRES model predictions, respectively).\\
The core C-burning temperature of each star is then shifted to hotter temperature as seen in Fig.~\ref{fig:crosstemp_rot} that gives the equivalent of Fig.~\ref{fig:crosstemp} for rotating models. We note, in particular, that the 10 M$_{\odot}$ now ignites carbon in core when rotation is considered (see also Fig.~\ref{fig:Tcrhoc10}). The observations of the impact of the resonance depending on the mass are similar to the ones discussed for the classical models, \textit{i.e.} a different impact of the resonance depending on the mass. Due to rotation, the C-burning temperature increases. In that context, the effective core C-burning temperatures of the 17~M$_\odot$ are close to the minimum difference of the CF88 and the HINRES rate localised by the green dotted vertical lines. Equally to the discussion of the classical models, stars lighter than $\approx$17~M$_\odot$ are affected at low temperatures (cold limit) and stars heavier than $\approx$17~M$_\odot$ are affected at high temperatures (hot limit). In addition, due to the temperature increase, the limit for core C-ignition is lowered in mass to the 10 M$_{\odot}$, that is now impacted by the resonance. However, the difference in the reaction rates reference does not affect significantly this ignition limit as discussed in Sect.~\ref{sect:massdep}. If C-ignition occurs, changing the rates impacts, nevertheless, the C-burning and chemical evolution that take place at higher density and temperature in the HIN/HINRES models as observed in Fig.~\ref{fig:Tcrhoc15}.
\subsection{Chemical abundances}
\label{sect:abond}
Changing the reaction rates in the models also has an impact on the chemical evolution predictions. From a classical model using the CF88 reference, the modification of the nuclear reaction rates, as well as the addition of the rotation-induced mixing along evolution, drive important changes in chemical abundances of the direct C-burning products, but also for the chemical structure itself.

\subsubsection{Classical versus rotating models}

\begin{figure*}[t]
         \center
         \includegraphics [trim = 13mm 8mm 3mm 5mm,clip,width=90mm]{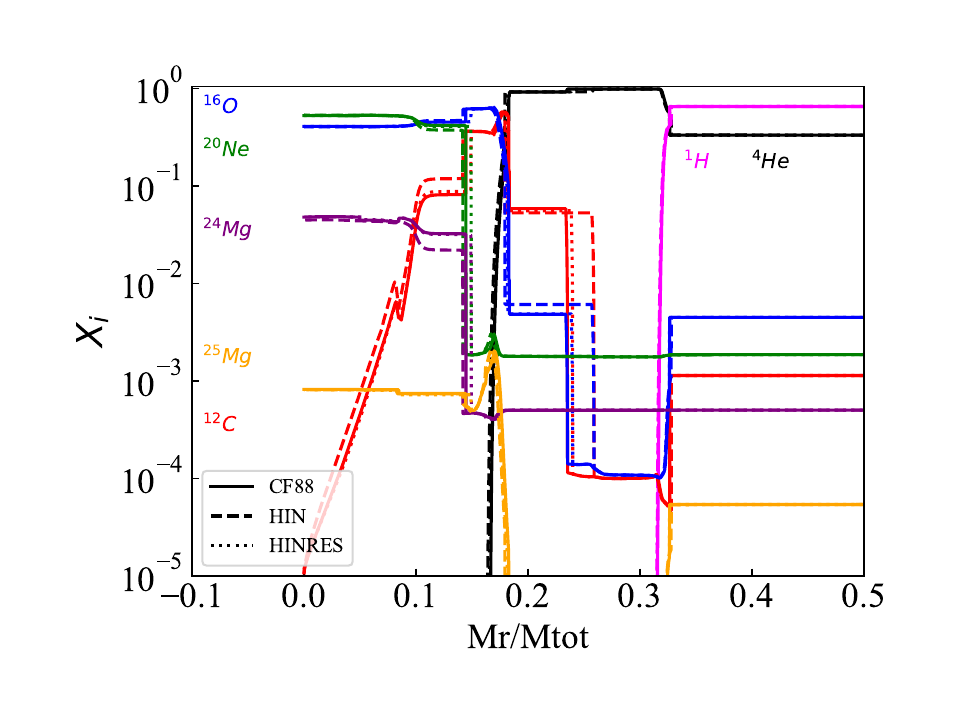}
         \includegraphics [trim = 13mm 8mm 3mm 5mm,clip,width=90mm]{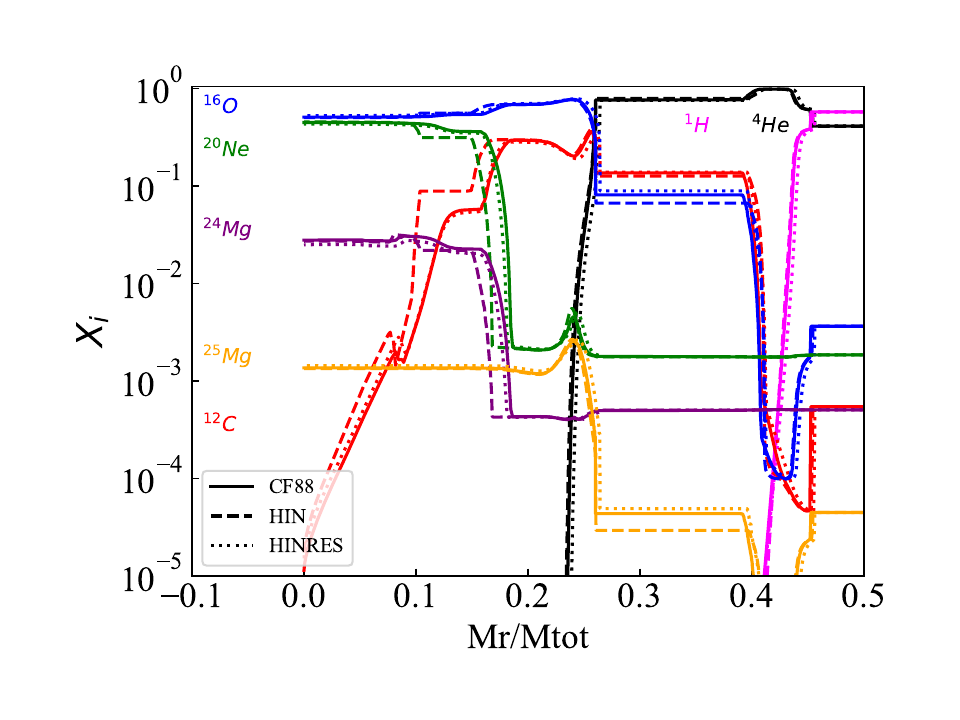}
         \includegraphics [trim = 13mm 8mm 3mm 5mm,clip,width=90mm]{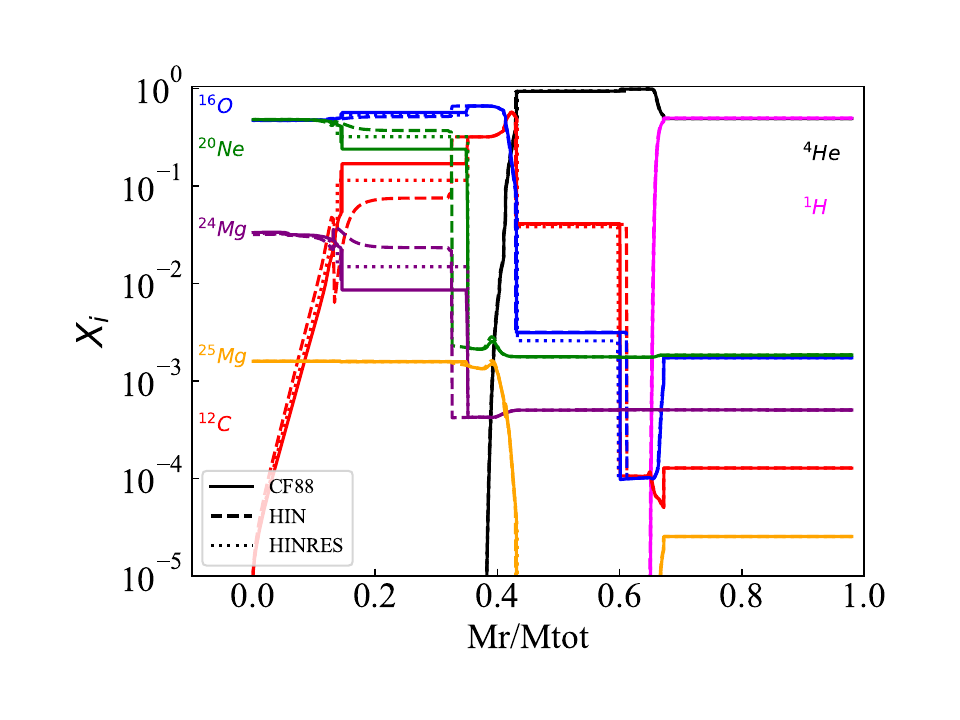}
         \includegraphics [trim = 13mm 8mm 3mm 5mm,clip,width=90mm]{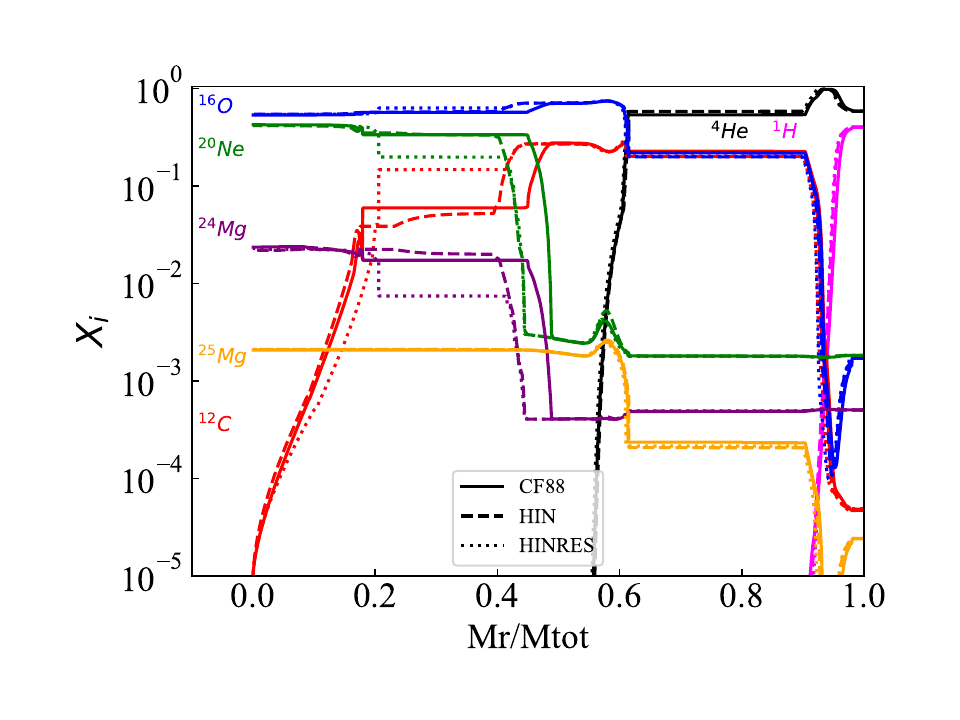}
         \includegraphics [trim = 13mm 8mm 3mm 5mm,clip,width=90mm]{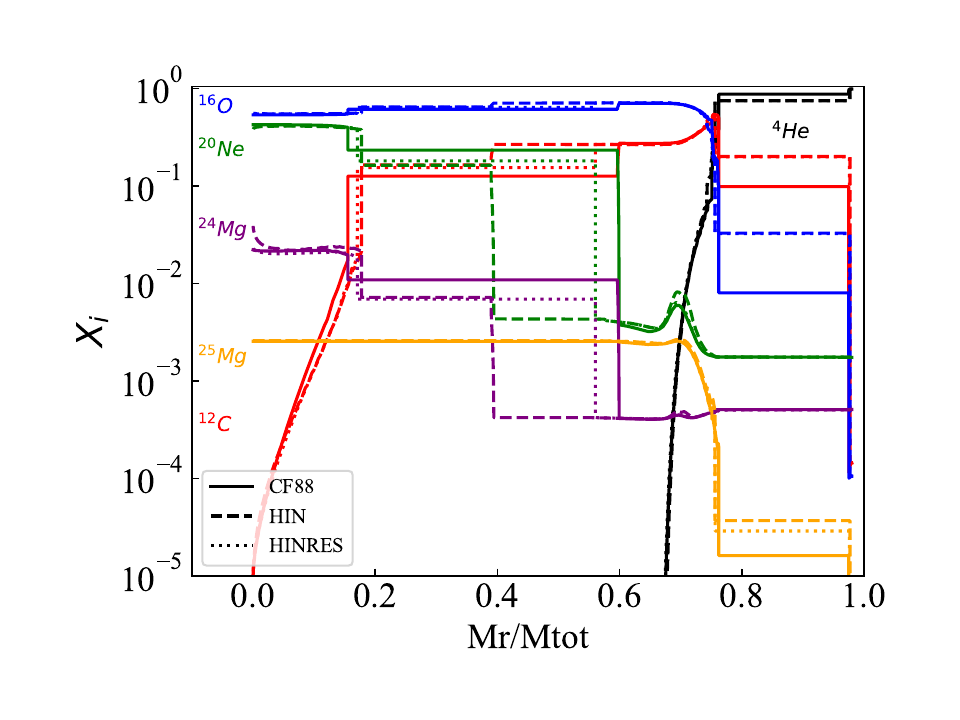}
         \includegraphics [trim = 13mm 8mm 3mm 5mm,clip,width=90mm]{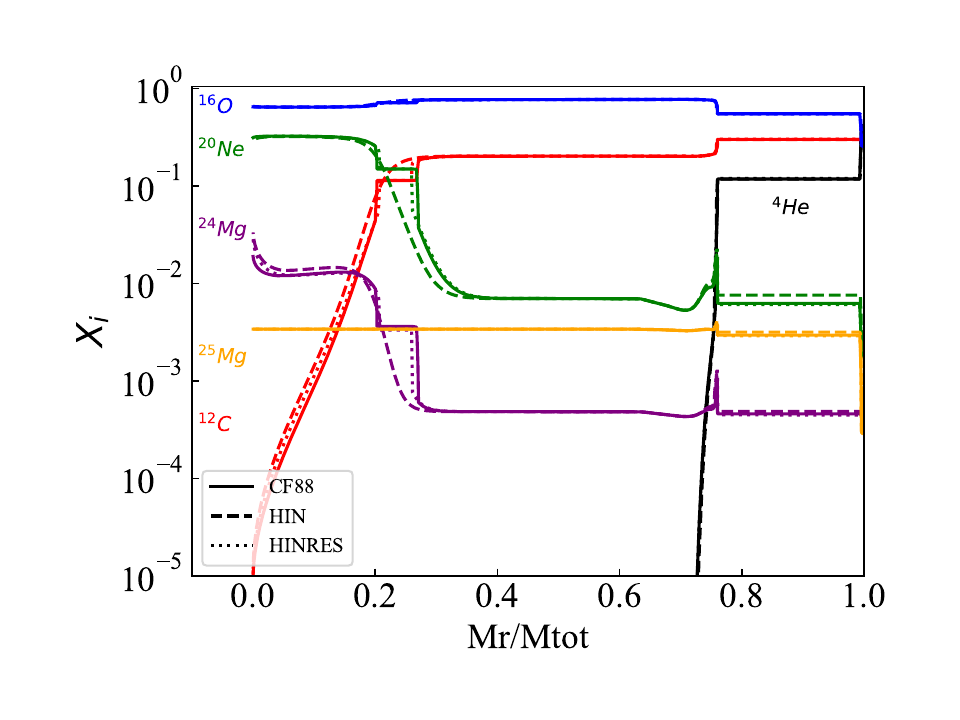}
         \caption{Profiles of abundances of $^{1}$H, $^{4}$He, $^{12}$C, $^{16}$O, $^{20}$Ne, $^{24}$Mg, and $^{25}$Mg (colour-coded), at the end of the C-burning phase for the 15, 20, and 30 M$_\odot$ stars models (from top to bottom) for different reactions rates (line-coded). Left: Classical models. Right: Rotating models. 
         }
         \label{fig:profmassive}
\end{figure*}

Figure~\ref{fig:profmassive} displays the abundance profiles for the 15, 20 and 30~M$_{\odot}$ stars at the end of the core C-burning phase for classical (left) and rotation (right) models. When carbon is exhausted in the core, an ONe core remains with Mg at a lower level. Similar to what was already widely shown in the literature comparing classical and rotating models \citep[\textit{e.g.}][]{2004A&A...425..649H,2009pfer.book.....M,2021FrASS...8...53E}, we observe a systematic increase of the core size and smoother chemical transitions for each mass, as usually observed when considering rotational-mixing or turbulent transport. For the 15 M$_{\odot}$ star, classical models predict Ne-dominated cores whereas the core becomes $^{16}$O-dominated at higher mass or when including rotation. As discussed for instance in \citet{2022A&A...659A.150D}, due to the complexity of the nuclear reactions involved, several contributions can be the cause of such a behaviour. A first explanation of the origin of such a core is the assumption in GENEC of merging all exit channels into the $\rm ^{12}C(^{12}C,\alpha)^{20}Ne$ reaction, averaging over the $Q$-values and branching ratios that sums the energies of the three channels but does not consider the chemical aspect. We then expect an overproduction of $\rm ^{20}Ne$ and an underproduction of $\rm ^{23}Na$ in addition to the direct contribution of $\rm ^{23}Na(p,\alpha)^{20}Ne$ and $\rm ^{16}O(\alpha,\gamma)^{20}Ne$ reactions that take place in parallel. This approximation correlates differently with the mass and with/without rotation giving birth to this difference of core abundances between the models. This aspect will be further discussed in Section~\ref{sect:nucleo} using a one-zone nucleosynthesis code taking into account the different exit channels and isotopes involved. A second part of the explanation comes from the mixing driven directly by rotation in the stellar core. At the end of the He-burning phase, when including rotation, the ratio of $\rm ^{12}C$ over $\rm ^{16}O$ is smaller. During the C-burning phase, $\rm ^{12}C$ is burned and the $\rm ^{20}Ne$ produced from the available $\rm ^{12}C$ fuel is then less abundant in rotating models. In parallel, the $\rm ^{16}O(\alpha,\gamma)^{20}Ne$ reaction tends to be less efficient at the higher temperatures predicted for rotating stars and/or more massive stars \citep{1989A&A...210..155M}, explaining the change on the core at the end of C-burning phase. The abundances of the other abundant elements are not significantly modified but the structure of abundances is changed. The size and physical nature (radiative or convective temperature gradient) of the core will be further discussed in Sect.~\ref{sect:quiescent}.
\subsubsection{Impact of nuclear reaction rates references}
\begin{figure}[h!]
         \center
         \includegraphics [trim = 13mm 8mm 3mm 5mm,clip,width=90mm]{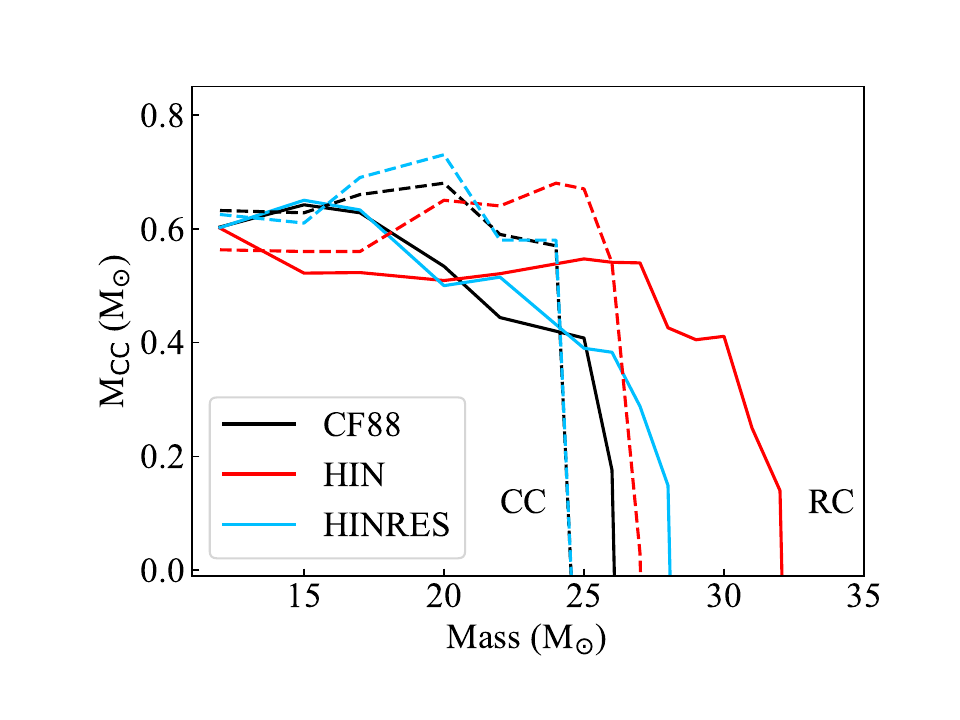}
         \caption{Maximum C-burning convective core size as a function of the initial mass for classical (solid lines) and rotation (dashed lines) models for the three reaction rates references (colour-coded). CC: convective core. RC: radiative core.}
         \label{fig:cc}
 \end{figure}

In Fig.~\ref{fig:profmassive}, we also compare the model predictions using the three different nuclear reaction rates references. For classical models, the change of rates does not affect much the abundances in the core, but in outer shells, and especially in the 30 M$_{\odot}$ model. In addition, the location of the structural transitions between the core and the successive shells are shifted depending on the nuclear rates as already showed by M22 for the 12 and 25 M$_{\odot}$ stars due to the effect on C-burning shells size. M22 noticed for instance that the transition between $^{12}$C-dominant and $^{20}$Ne-dominant regions are shifted outwards (inwards) for the 25 (12) M$_{\odot}$ HIN model compared to the CF88/HINRES models. This is confirmed by our results with a first predicted transition taking place close to the classical 22 M$_{\odot}$ and a second one beyond the classical 27 M$_{\odot}$. In other words, for stars less massive than 22 M$_{\odot}$, the $^{12}$C-dominant/$^{20}$Ne-dominant transition of the CF88/HINRES models compared to the HIN models is shifted inwards, between 22 and 27 M$_{\odot}$, it is shifted outward, and beyond it is shifted inward again as the formation of the convective core itself is then impacted. At a lower level, some differences are also observed between the CF88 and HINRES predictions, and especially for the classical 30 M$_{\odot}$ and rotating 20 M$_{\odot}$, where the Ne/C transition is shifted inward for the HINRES model. \\ It shows the changes of structure and different extension of the convective zones (core and shells) depending on the mass and on the nuclear reference. This is also the conclusion from Fig.~\ref{fig:cc} that displays the maximum convective core mass as a function of the initial mass for each model during the C-burning phase. It shows that HIN models will exhibit lighter convective cores between 12 and 20 M$_{\odot}$, and heavier ones beyond 20 M$_{\odot}$. Models computed with the HIN reference predict the formation of a convective core up to higher initial masses than CF88/HINRES references. In addition, the transition from a convective core to a radiative core is observed in the chemical profiles of the classical 30 M$_{\odot}$ between the different nuclear cases (see Sect.~\label{sect:quiescent} and Fig.~\ref{fig:cc}).\\ The chemical structure is changed as well in the inner regions of the star for $\rm^{20}Ne$ and $\rm^{24}Mg$ while not affecting much $\rm^{25}Mg$. The rotational-mixing tends to smooth these differences. These chemical differences may drive the neutron seed production, and consequently, impact the s-process efficiency during this phase and the following ones (see Sect.~\ref{sect:nucleo}). Finally, surface abundances are not significantly impacted while it is mainly the abundance structure of the star that is affected when changing the rates, as a result of different temperature and size of the convective C-burning core and C-burning shells when changing the nuclear rates. Surface abundances are mainly the result of the rotation-induced mixing along evolution.

\section{Stellar critical mass limits}
\label{sect:massdep}

As introduced in Sect.~\ref{section:introduction}, the C-ignition and the C-burning phase are keys that will determine the end-of-life of stars. The change of core physical conditions due to the change of nuclear reaction rates is then especially interesting in the context of the stellar mass limits \citep[\textit{e.g.}][]{1989A&A...210..155M,2009pfer.book.....M}. In the following section, we will mainly focus, first, on the physical nature (convective or radiative temperature gradient) of the core depending on the mass and rates at about 20-30 M$_{\odot}$ \citep[\textit{e.g.}][]{2013ApJ...762...31P,2021ApJ...916...79C}, and second, on the so-called M$_{\rm up}$ and M$\rm _{mas}$ critical limits at the transition between intermediate and massive stars \citep[\textit{e.g.}][and reference therein]{2006A&A...448..717S,straniero2016,2022A&A...659A.150D,2023MNRAS.525.3216C,Limongi2024ApJS}.
 
\subsection{\TDD{Formation of a convective C-burning core}}
\label{sect:quiescent}
In addition to the core size, the effects of rotation and the change of nuclear reaction rates impact \TDD{the nuclear energy flux and by this way} the temperature gradient \TDD{and the formation, or not, of a convective C-burning core \citep[see also description in Sect.~4 by][]{2021ApJ...916...79C}}. Figure~\ref{fig:cc} shows the mass limit between stars where a convective zone is formed during the core C-burning phase and stars that remain with a radiative core, for classical and rotating models (also reported in Tab.~\ref{tab:lifetime} for the present work and several references). For this analysis, we computed additional models at intermediate masses of 24, 26, 27, 28, 29, and 32 M$_{\odot}$. \\
\newline
 The main effect of decreasing the rates on the structure of stars is to shift this limit to higher mass and change the size/mass of the convective core if formed. The mass limit to form a convective core was previously determined by \textit{e.g.} \citet{2004A&A...425..649H} at less than about 22 M$_{\odot}$, between 20 and 25 M$_{\odot}$ by \citet{2012MNRAS.420.3047B}, and less than 25 M$_{\odot}$ by \citet{2021ApJ...916...79C}, using the CF88 reference in classical models. In the present work, we determine this limit at less than 27 M$_{\odot}$, using the CF88 reference in a classical model at solar metallicity. When considering the new references HIN and HINRES, this limit is shifted to higher masses at less than 32 M$_{\odot}$ and less than 29 M$_{\odot}$, for HIN and HINRES models, respectively. This shift is expected from the higher temperatures reached by these two models compared to the CF88 (see Fig.~\ref{fig:rhot}). Indeed, the formation of the convective core is achieved when the nuclear energy from C-burning E$\rm _{nucl.C}$ is higher than the energy lost by escaping neutrinos E$_{\nu}$ \TDD{and when E$\rm _{nucl.C}$ is high enough that the radiative gradient overcomes the adiabatic one}. E$_{\nu}$ is being higher at hotter temperatures and higher densities reached in the core when using the lower rates of the HIN/HINRES models, finally leading to the shift to higher masses of the convective core formation. This relative shift is consistent with previous studies that explored the other way around and showed that the use of higher nuclear rates compared to CF88 generally results of a shift of this limit to smaller masses due to colder and less dense C-burning cores \citep{2012MNRAS.420.3047B,2013ApJ...762...31P,2021ApJ...916...79C}. In particular, \citet{2021ApJ...916...79C} determined, using the results from \citet{2018Natur.557..687T}\footnote{Trojan horse indirect determination of the cross sections for the $\rm ^{12}C+^{12}\!C$ fusion reaction.}, this limit at less than 21 M$_{\odot}$ due to the very high measured cross section values. When formed, the size of the core is also impacted when using different rates. We predict a convective core of $\approx$ 0.65 M$_{\odot}$ and 0.53 M$_{\odot}$ for the 15 and 20 M$_{\odot}$ stars, respectively, using CF88. Changing the reference to HIN rates lead to smaller core sizes between 15 and 20 M$_{\odot}$ ($\approx$ 0.50 M$_{\odot}$ for both the 15 and 20 M$_{\odot}$), while giving similar results using HINRES. This is in good agreement with the result from \citet{2021ApJ...916...79C} both in terms of absolute value of the core using the CF88 reference and in terms of relative effects of the change of the rates: using the higher THM nuclear rates they predict larger convective cores ($\approx$ 1.00 M$_{\odot}$ for the 15 M$_{\odot}$) between 12 and 20 M$_{\odot}$. According to the analysis by \citet{2021ApJ...916...79C}, the compactness of the core is then expected to increase in HIN/HINRES models and may affect the final fate and explosibility of these massive stars as also described \textit{e.g.} by \citet{2016ApJ...818..124E,2020MNRAS.499.2803P}. We stress that the choice of the nuclear rates reference is consequently primordial for model predictions as it lead to significantly different results for the characteristics of the core in this mass range from using THM rates or HIN rates, the two extreme cases. It then impacts significantly the chemical evolution and structure, as well as the end of life of stars, already in classical models. The addition of rotation, changing the core parameters, drives a global shift to lower masses for the convective core formation with a decrease at less than 25, 28, and 25 M$_{\odot}$ for CF88, HIN, and HINRES models, respectively. On the other hand, rotation drives the formation of larger convective cores between M$_{\rm ini}$ $\approx$ 15-17 and $\approx$ 25 M$_{\odot}$ impacting the global mixing of chemicals (see also Fig.~\ref{fig:profmassive}).\\
\newline
The choice of the nuclear reaction rates is shown to be important for the 12-30 M$_{\odot}$ mass range impacting the formation and size of the convective core during C-burning phase, and consequently, the chemical evolution and fate of stars. Several nuclear references are available in recent literature and the model predictions are highly dependent on the choice of the reference, and consequently, should take into account the nuclear uncertainties. Also, rotation has a strong contribution to the evolution of the stellar core and should be considered for more realistic predictions and results in general. On the other hand, it has always to be reminded that the prescriptions used to describe rotation in 1D stellar evolution codes like GENEC are approximations from 3D hydrodynamical simulations and that several prescriptions are available potentially impacting the transport efficiency \citep[\textit{e.g.}][]{2013LNP...865....3M,2021A&A...646A..48D,Nandal2023}. Important uncertainties then remain and should be explored.  

\subsection{Off-centre C-ignition}
\label{sect:offcentre}

\begin{figure}[t]
         \center
         \includegraphics [trim = 13mm 8mm 3mm 5mm,clip,width=90mm]{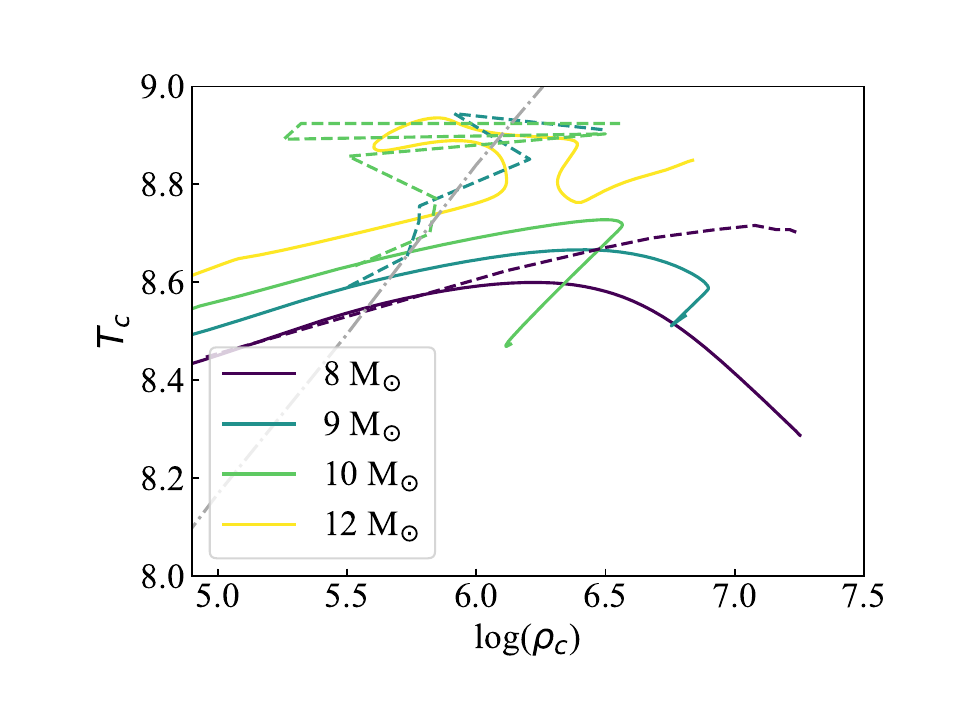}
         \caption{Evolution of T$_{\rm C}$ and $\rm \rho_{C}$ for classical models of 8, 9, 10, and 12~M$_{\odot}$ stars with the CF88 reference. Dashed lines indicate the maximum off-centre temperatures and corresponding density (except for the 12 M$_{\odot}$ star for which the maximum temperature corresponds to the core). The grey dotted-dashed line shows the line of degeneracy.}
         \label{fig:off}
 \end{figure}

\begin{figure}[t]
         \center
         \includegraphics [trim = 13mm 8mm 3mm 5mm,clip,width=90mm]{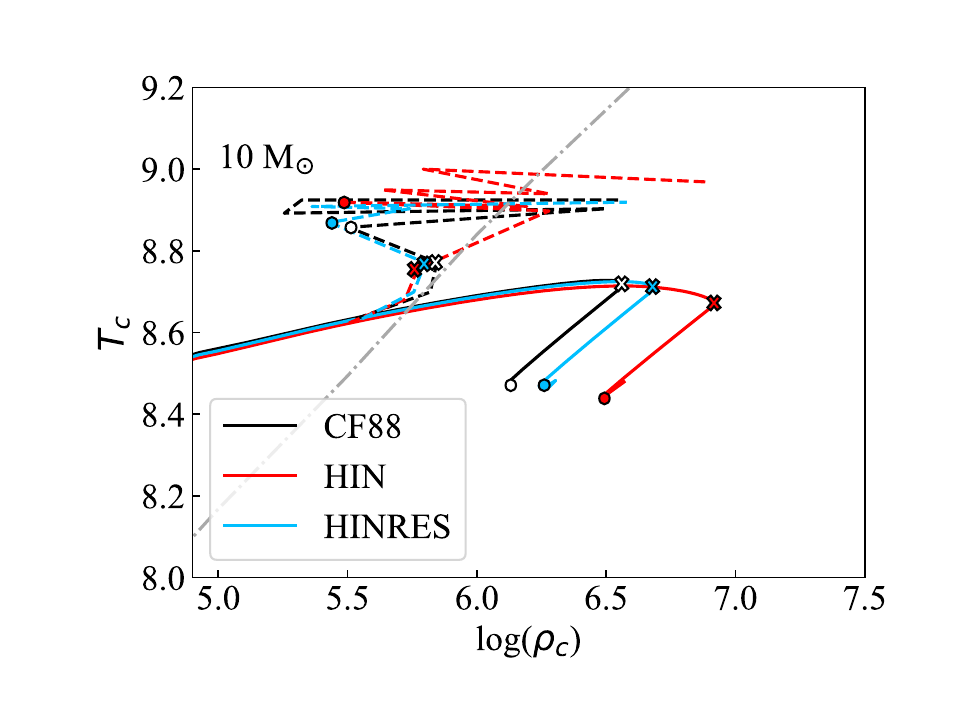}
         \caption{Evolution of T$_{\rm C}$ and $\rm \rho_{C}$ for classical models of a 10 M$_{\odot}$ star with CF88, HIN, HINRES references (colour-coded). Dashed lines indicate the maximum off-centre temperatures and associated density. Crosses and dots shows the off-centre C-ignition and beginning of \TDD{flashes} for each model (colour-coded). The grey dotted-dashed line shows the line of degeneracy.}
         \label{fig:offcomp}
 \end{figure}

\begin{figure}[t]
         \center
         \includegraphics [trim = 13mm 8mm 3mm 5mm,clip,width=90mm]{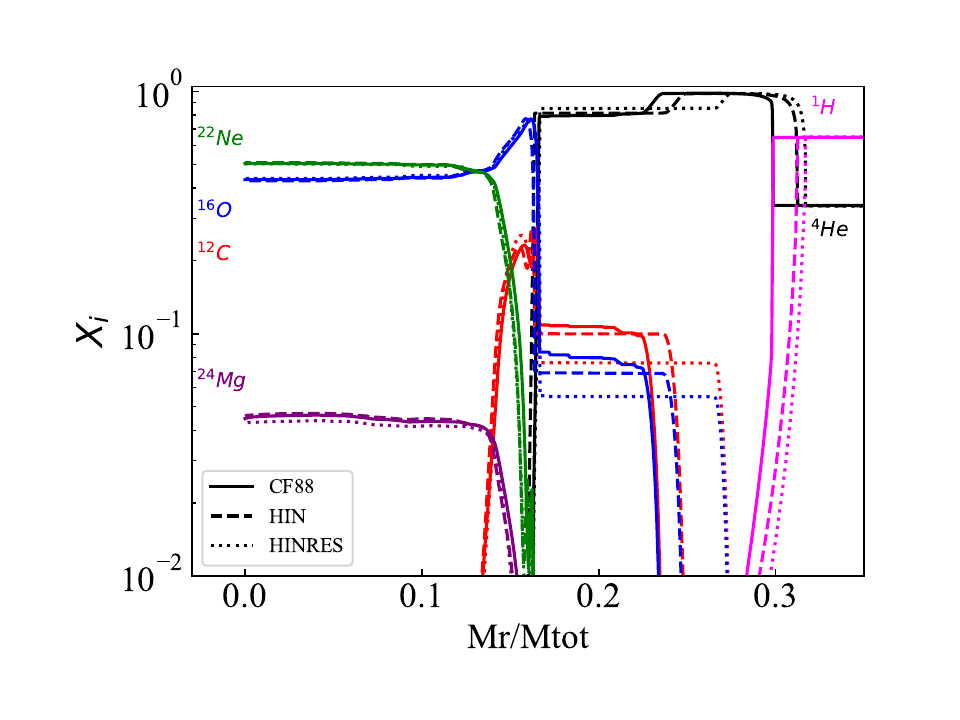}
         \caption{Profiles of $^{1}$H, $^{4}$He, $^{12}$C, $^{16}$O, $^{20}$Ne, and $^{24}$Mg as a function of $\rm Mr/M_{tot}$ at the end of the core C-burning for a rotating 10 M$_{\odot}$, and for the three rates references (CF88: full line, HIN: dashed line, HINRES: dotted line).}
         \label{fig:profM10}
 \end{figure}

In the case of the lowest mass stars of our grid (\textit{\textit{i.e.}} $\rm \le 10 M_{\odot}$), our models predict no C-ignition in the centre, that is degenerate (at the exception to the 10 M$_{\odot}$ with rotation), but rather off-centre ($0 < \rm Mr/M_{tot} \le 0.1$), that is above the degeneracy limit, at higher temperature.
Figure~\ref{fig:off} shows the evolution of the \TDD{central} temperature and \TDD{central} density (full lines) along with the maximum off-centre temperature and associated density (dashed lines) for classical models of 8, 9, and 10 M$_{\odot}$ using CF88 reference, as well as the 12 M$_{\odot}$ star that ignites carbon in its core for comparison.
Carbon burning ignites off-centre for the classical 9 and 10 M$_{\odot}$ but not for the classical 8 M$_{\odot}$, that stays too cold. As discussed by \textit{e.g.} \citet{1989A&A...210..155M}, the point of the maximum temperature is shifted outwards the centre, due to the neutrino loss, and can reach the condition for C-ignition. In Fig.~\ref{fig:off}, the ignition takes place at \TDD{$\approx$~0.07 $\rm Mr/M_{star}$ and $\approx$~0.02 $\rm Mr/M_{star}$ for the 9 and 10 M$_{\odot}$, respectively}. Then, a flame arises and may propagate to the centre, allowing the usual massive star evolution \citep[see for instance the series of paper][]{2006A&A...448..717S,2007A&A...476..893S,2010A&A...512A..10S}. In that sense, it can be seen as the carbon flash, analogously to the well-known helium flash of low mass stars. It is in this range of masses, between M$\rm _{up}$ and M$\rm _{mas}$, that takes place the so-called SAGB stars that will end their life as electron capture supernovae, ONe white dwarfs or in a few cases as hybrid CO-Ne white dwarfs if the flame does not reach the core \citep[e.g.][]{2010MNRAS.401.1453D,2014MNRAS.437..195D,2015MNRAS.446.2599D,2015ApJ...807..184F,Limongi2024ApJS}.\\
The limit of the off-centre C-ignition is then impacted by the nuclear reactions as the core C-ignition and, consequently, it also constrains the M$\rm _{up}$ limit between AGB and SAGB stars, as well as the evolution of super-AGB stars and the formation of the massive white dwarfs in this mass range \citep[\textit{e.g.}][]{2018MNRAS.480.1547L,2022A&A...659A.150D,2023arXiv230300060D} below M$\rm _{mas}$. The limit M$\rm _{up}$ is usually defined to be between 7.5 and 9 M$_{\odot}$ depending on the stellar evolution codes and input physics considered \citep[\textit{e.g.}][]{2007A&A...476..893S,2008ApJ...675..614P,straniero2016}, with a CO core of 1.1-1.3 M$_{\odot}$ before C-ignition. In the present study, our classical models using the CF88 reference predict a CO core of 0.80, 1.10, and 1.25 M$_{\odot}$, and our rotating models a CO core of 1.15, 1.17, and 1.70 M$_{\odot}$ for the 8, 9, and 10 M$_{\odot}$, respectively. The addition of rotation drives an increase of the core size for each mass allowing the off-centre C-ignition for the 8 M$_{\odot}$ and the quiescent ignition for the rotating 10 M$_{\odot}$ (see Fig~\ref{fig:Tcrhoc10}). These results are in agreement with previous results of the literature and mark the limit M$\rm _{up}$ between 9 and 10 M$_{\odot}$ for classical models and 8 and 9 M$_{\odot}$ for rotating models, using the CF88 reference, also in agreement with the critical mass for C-ignition close to the value of 1.05 M$_{\odot}$ \citep{1989A&A...210..155M,2007A&A...476..893S,2009pfer.book.....M,Doherty2017,2022A&A...659A.150D}.\\  \TDD{Similarly to M$\rm_{up}$, we determine the M$\rm _{mas}$ limit beyond which Ne is ignited\footnote{In order to explore this mass limit we extended our grid with two additional models of 11 M$_{\odot}$ with or without rotation and CF88 reference.}. From the non-rotating models, \citet{Doherty2017} obtained a value of M$\rm _{mas}=11$ M$_{\odot}$ with a minimum ONe core mass of 1.37 M$_{\odot}$ for Ne-ignition \citep{Nomoto1984}, while \citet{Limongi2024ApJS} recently determined a smaller value of M$\rm _{mas}=9.22$ M$_{\odot}$ with a minimum ONe core mass of 1.349 M$_{\odot}$. In addition, \citet{Limongi2024ApJS} also showed that the CO core mass was more relevant and robust and determined a minimum CO core mass for Ne-ignition at 1.363 M$_{\odot}$. From our models we obtained: without rotation, a M$\rm _{mas}$ limit between 11 and 12 M$_{\odot}$ with an ONe core of about 1.48 M$_{\odot}$ and a CO core of 1.58 M$_{\odot}$ for the 12 M$_{\odot}$ at Ne-ignition; with rotation, we determined the M$\rm _{mas}$ limit between 9 and 10 M$_{\odot}$ with an ONe core of 1.45 and a CO core of 1.60 M$_{\odot}$ for the rotating 10 M$_{\odot}$ at Ne-ignition. Our results from non-rotating models are in good agreement with those by \citet{Doherty2017} and \citet{Limongi2024ApJS}, although there is a small shift exists with the latter. We note that such a small difference is expected as the structure is sensitive to model physics inputs such as the treatment of the overshoot.} \\
 \newline
 Figure~\ref{fig:offcomp} shows the comparison using the three different nuclear reaction rates references for the classical 10 M$_{\odot}$ star, for both core and off-centre conditions. The C-ignition takes place in each case (colour-coded crosses) \TDD{and should be followed by} the arising of one or several carbon flash(es) (colour-coded dots) \TDD{and} the flame propagation. We note that the GENEC code cannot follow the carbon flash(es) and the flame propagation. We, consequently, focus on the analysis of the off-centre C-ignition only while the flame propagation will be studied in a future work.\\ The decrease of the nuclear rates results in a shift of the C-ignition to higher temperatures and densities \citep[\textit{e.g.}][]{2007PhRvC..76c5802G,2013ApJ...762...31P} for the HINRES and HIN models compared to the CF88 model, as observed for non-degenerate core C-burning of more massive stars. These differences, although small, can be at the origin of different flame propagation and chemical abundances at the end of the C-burning phase. For instance, \citet{Chen2014} showed that hindrance could lead to a slight increase of M$\rm _{up}$. However, the change of nuclear reaction rates in our models does not modify the limit between core ignition and off-centre C-ignition, or the M$\rm _{up}$ limit regarding the mass step of 1 $M_{\odot}$ we considered in our grid. It is interesting to notice as \citet{straniero2016} found that a hypothetical narrow resonance at $\approx$ 1.4 MeV would reduce the M$\rm _{up}$ limit of about 2 M$_{\odot}$ down to 5.8 M$_{\odot}$. At the opposite, HIN/HINRES models show no significant (> 1 M$_{\odot}$) increase of this limit. We note that the classical and rotating 9 M$_{\odot}$ (not shown) behave like the classical 10 M$_{\odot}$ but at lower temperatures, and the rotating 8 M$_{\odot}$ (not shown) shows only traces of C-ignition in an off-centre layer. \\
 \newline
With the addition of rotation, the 10 M$_{\odot}$ star is now igniting carbon. Figure~\ref{fig:profM10} shows its abundance profiles at the end of the core C-burning phase for $^{1}$H, $^{4}$He, $^{12}$C, $^{16}$O, and $^{20}$Ne, and $^{24}$Mg, with the three nuclear references. The addition of rotation-induced mixing is impacting the structure and chemicals, and changes the fate of stars for a same initial mass. The addition of rotation affects the stellar evolution before reaching the C-ignition and then changes the characteristics of the core during the C-burning phase. With the exception to the rotating 10 M$_{\odot}$, the lower mass stars may end their life as massive hybrid CO-Ne white dwarfs, or potentially, depending on the internal processes taking place in the stars, as ONe white dwarfs \citep[\textit{e.g.}][]{2018MNRAS.480.1547L,2022A&A...659A.150D}. In particular, \citet[][]{2009A&A...497..463S,2022A&A...659A.150D} showed that convective boundary mixing, a strong overshoot or thermohaline mixing during the C-burning phase is leading to the halt of the carbon flame and is at the origin of the hybrid CO-Ne structure. The mass range for the formation of such a hybrid core is, however, small. \citet{2015MNRAS.446.2599D} gives a range of 0.3 M$_{\odot}$ below the pure ONe white dwarfs and M$\rm _{up}$ $\approx$ 7 $M_{\odot}$, and \citet{2015ApJ...807..184F,Limongi2024ApJS} give a range of 1 M$_{\odot}$ below the pure ONe white dwarfs and M$\rm _{up} \approx 8$ M$_{\odot}$. A refined grid of models between 8 and 10 M$_{\odot}$ should be realised in order re-estimate these values regarding the new reaction rates and their impact on the flame propagation, as mentioned below. \\
\newline
At the end of the C-burning phase of \TDD{the rotating 10 M$_{\odot}$}, our models predict that $^{20}$Ne is the most abundant element in the ONe shell \TDD{(the production of $^{23}$Na is not followed during the C-burning phase and its abundance stays at a very low level of} X($\rm ^{23}Na) < 10^{-2}$, not appearing in Fig.~\ref{fig:profM10}) as observed by \citet[\textit{e.g.}][]{2018MNRAS.480.1547L}, but in contrary to results from \textit{e.g.} \citet{2006A&A...448..717S} where $^{16}$O is the most abundant. As described in the previous section, it is partly the result of the nuclear reaction network used in this work where we considered all the channels on the energetic point of view but only the reaction $^{12}$C($^{12}$C,$\alpha$)$^{20}$Ne for nucleosynthesis and not the reaction $^{12}$C($^{12}$C,p)$^{23}$Na. It was already observed by \citet[][see for instance their Figure 7]{2022A&A...659A.150D} who add that the produced $\alpha$ particles participate as well to the high abundance of $^{20}$Ne through the reaction $^{16}$O($\alpha$,$\gamma$)$^{20}$Ne. \\ It is clear that a larger reaction network is required in order to explore further the chemical evolution of these stars, as well as the consequences for the white-dwarf evolution itself. In addition, transport processes like rotation-induced mixing in the present work and/or convective boundary mixing as explored in \citet[][]{2022A&A...659A.150D} should be considered as well in order to achieve more realistic chemical predictions and final fate of stars. When rotation is included, the change of nuclear reaction rates is not significantly altering the results for the 10 M$_{\odot}$ but it will affect the propagation of the flame in the 9 M$_{\odot}$ for instance, with especially different temperatures of the flame as well as shifted chemical transitions. These results are in agreement with \citet{2023arXiv230300060D} who, in particular, mentioned as well that it will lead to different pulsation patterns for these white dwarfs.

\section{Nucleosynthesis}
\label{sect:nucleo}

The modification of the nuclear reaction rates is known to affect the stellar nucleosynthesis and final stellar abundances, with especially the neutron capture processes \citep[\textit{e.g.}][]{2012MNRAS.420.3047B,2013ApJ...762...31P,2016MNRAS.456.1803F,2018A&A...618A.133C}. As described in Sects.~\ref{GENEC} and \ref{sect:abond}, we do not include a large nuclear reaction network in our stellar evolution models. In order to further explore the effect on nucleosynthesis, in particular for s-process elements, we use in a second step a one-zone code \citep{2016A&A...593A..36C} that allows to consider large networks (and especially including the three exit channels of the carbon fusion). We recall that such a code involves two important limitations: the computation is done only for the centre and does not consider any transport processes \TDD{like, especially, convective mixing and rotational mixing}. However, as described in Sect.~\ref{sect:nucl}, it has been shown that such a code is reliable to explore nucleosynthesis in both non-rotating and rotating models if we assume that the C-burning timescale is very short compared to the \TDD{rotation} mixing timescale, which is verified for the stars of our grid as shown in Tab.~\ref{tab:lifetime}, and if we consider only the core C-burning phase \citep[see also similar studies in the case of He-burning phase by e.g.][]{Frost-Schenk22,Williams2022}. Here we considered a network of 736 isotopes which is the same as the one used to follow s-process nucleosynthesis in \cite{2016MNRAS.456.1803F} and \cite{2018A&A...618A.133C}. In M22, similar nucleosynthesis tests were performed but with a network of 1454 isotopes. We checked that the reduced and large networks lead to very similar results (the abundance differences are less than $0.01$~dex). 
We extracted the central temperature and density evolution tracks during the C-burning phase for each of the considered GENEC models and computed the resulting centre nucleosynthesis during the core C-burning phase as done in M22. We carried out the computation for four cases: 1) CF88 reference for the three channels; 2) HIN reference for $\alpha$ and proton channels plus neutron channel by \citet{2015PhRvL.114y1102B}; 3) HINRES reference for $\alpha$ and protons channels plus neutron channel by \citet{2015PhRvL.114y1102B}; and 4) CF88 reference for $\alpha$ and proton channels plus neutron channel by \citet{2015PhRvL.114y1102B}. \\

\subsection{Evolutionary path choice}

\begin{figure*}[t]
    \centering
    \includegraphics[width=90mm]{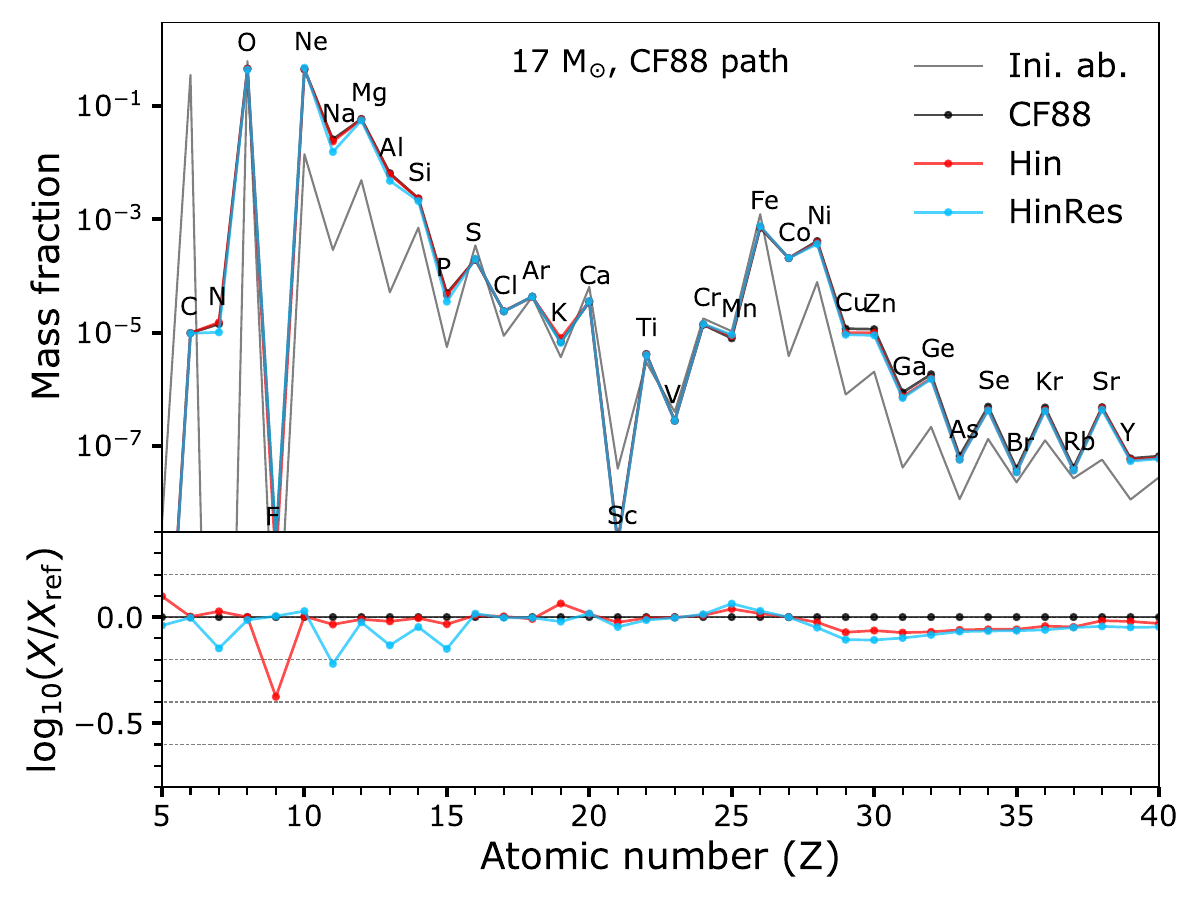}
    \includegraphics[width=90mm]{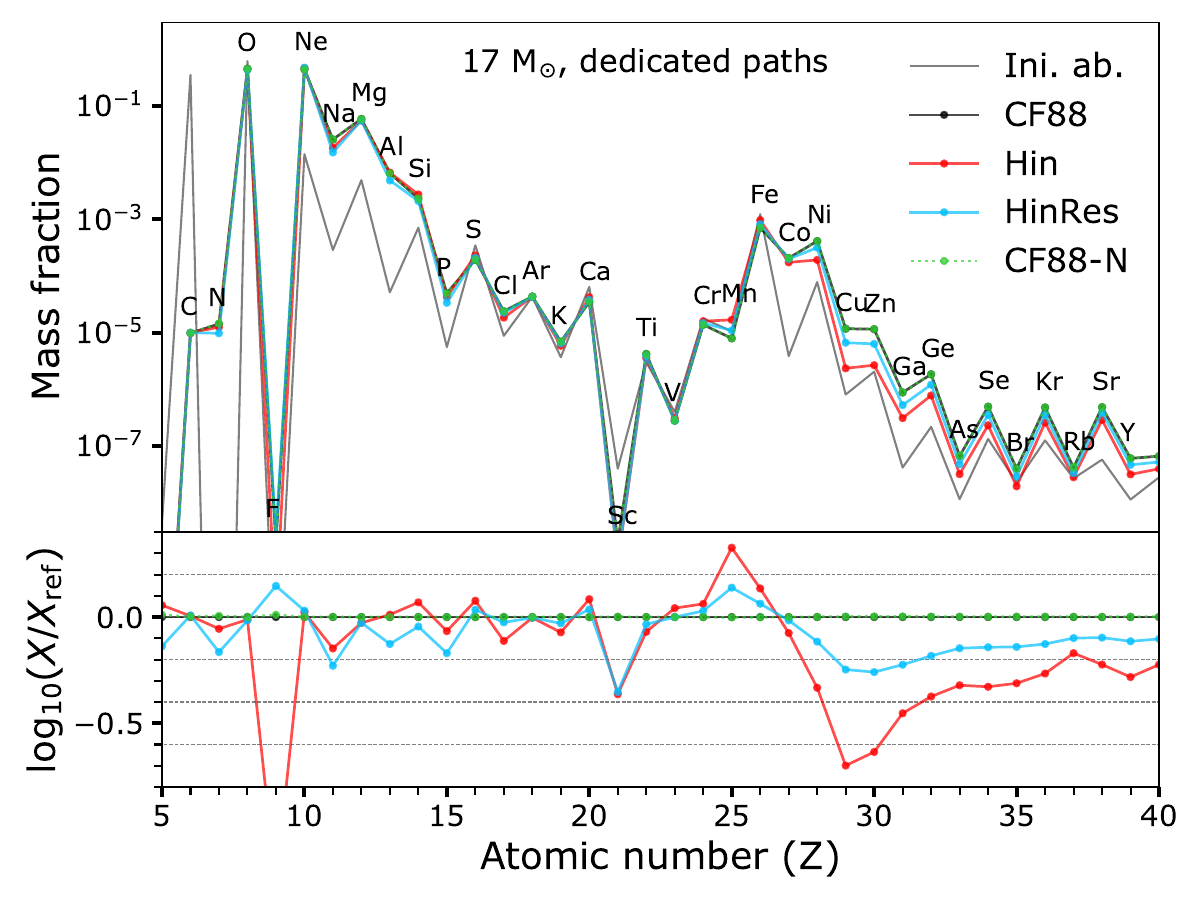}
    \caption{Mass fractions at the end of the core C-burning phase of a 17 M$_{\odot}$ star for three (or four) nuclear rates references \citep[colour-coded, CF88-N refers to CF88 + neutron channel by][]{2015PhRvL.114y1102B}. 
    The grey line refers to the initial abundances, extracted from the CF88 GENEC model at core C-ignition.
    The bottom panels show the abundances normalised to the ones of the CF88 model. Left: nucleosynthesis using the same ($\rho, T$) path from CF88 model. Right: nucleosynthesis using the consistent ($\rho, T$) paths from each GENEC model. 
    }
    \label{fig:nucleo_multi}
\end{figure*}

We first explored the impact of the thermodynamics conditions on our results. An evolutionary path consists of the temperature and density trajectories extracted from the complete stellar models and corresponding to the C-burning phase for the present study. Each evolutionary path is used as an input in the one-zone code, in contrast with the study by M22 where a unique evolutionary path from the CF88 model of a 25 M$_{\odot}$ was used, as a first approximation, to compute the three different nucleosynthesis scenarios. However, we saw that the change of nuclear reaction rates lead to different temperature and density paths during the C-burning phase (see Figs.~\ref{fig:rhot}, \ref{fig:rhot_bis} and \ref{fig:Tcrhoc10}).\\ Figure \ref{fig:nucleo_multi} shows the elemental abundances at the end of the one-zone C-burning phase, corresponding to a classical 17 M$_\odot$ model star, computed using only the CF88 path (left panel) and computed using the consistent paths for each case (right panel).
When using the consistent paths the HIN and HINRES models predict both a smaller abundance for Na (underproduced by about 0.2~dex) than the CF88 model whereas HIN model is similar to CF88 using CF88 path. Ne and Mg, the other main products of C-burning, are almost not affected in every case.\\ 
Concerning the heavy elements, there is a modest production of first-peak s-elements in all models. The three main neutron sources are $\rm ^{13}$C$\rm (\alpha,n)^{16}O$, $\rm ^{17}$O$\rm (\alpha,n)^{20}Ne$ and $\rm ^{22}$Ne$ \rm (\alpha,n)^{25}Mg$. We checked that  reducing these rates by a given factor (e.g. 1000) weaken the s-process signature, which becomes almost null if these rates are sufficiently small. The s-process production for the three models is similar when using the CF88 path (left panel) while differences of up to 0.7 dex (for $Z \simeq 30$) are noticed if adopting the consistent paths (right panel).
In the latter case, the s-process is more efficient in the CF88 model (black), followed by the HinRes (blue) and then the Hin model (red).
The maximal neutron densities are similar in these models, at about $2-4 \times 10^{8}$~cm$^{-3}$. 
The different s-process productions can then be understood through the different neutron exposures. 
In these one-zone simulations, the C-burning lifetimes of the CF88, Hin and HinRes models are 1480, 740 and 1450~yrs\footnote{In our one-zone model, at each time-step, the central $^{12}$C abundance extracted from complete stellar models is used to fix the corresponding temperature and density. The time evolution is not based on full stellar models, explaining the different C-burning lifetimes compared to full models (cf. Tab.~\ref{tab:lifetime}). Nevertheless, the duration of the C-burning phase follows the same trends when adopting different nuclear references.}, respectively, with total neutron exposure\footnote{The neutron exposure $\tau$ is computed as 
$\tau \, = \int  N_{\rm n}(t) \, v_{\rm T}(t) \, \text{d} t$ where $N_{\rm n}$ is the neutron density and $v_{\rm T} = \sqrt{ \, 2 \, k_B \, T(t) /m_{\rm n}}$ the neutron thermal velocity with $k_B$ the Boltzmann constant, $T(t)$ the temperature at time $t$, and $m_n$ the neutron mass.} of 0.19, 0.13 and 0.16~mbarn$^{-1}$.
Ultimately, the longer the C-burning lifetime, the higher the neutron exposure and the greater the number of light s-elements produced. 
\\ 
We conclude that using the consistent path for computing the nucleosynthesis has a non-negligible impact.\\
\newline 
On the right panel of Fig.~\ref{fig:nucleo_multi} we also consider a fourth case exploring the impact of the new rates determined by \citet{2015PhRvL.114y1102B} for the neutron channel while keeping the CF88 reference for $\alpha$ and proton exit channels (green pattern). Due to its comparably high $Q$-value, and hence energy threshold, this channel does not involve significant changes in the stellar structure of the models in the presented mass range, but could impact the s-process efficiency increasing the neutron density at high temperatures. However, we see that it does not affect the nucleosynthesis of the 17 M$_{\odot}$ star during the core C-burning phase (changes of $\approx$ 0.01 dex at most). Additional tests, for higher masses or during the shell C-burning, at higher temperature might be more relevant concerning this specific channel \citep[e.g.][]{2013ApJ...762...31P} and will be explored in a future study. 

\subsection{Nucleosynthesis with rotational-induced mixing}

\begin{figure*}[t]
         \center
         \includegraphics [width=90mm]{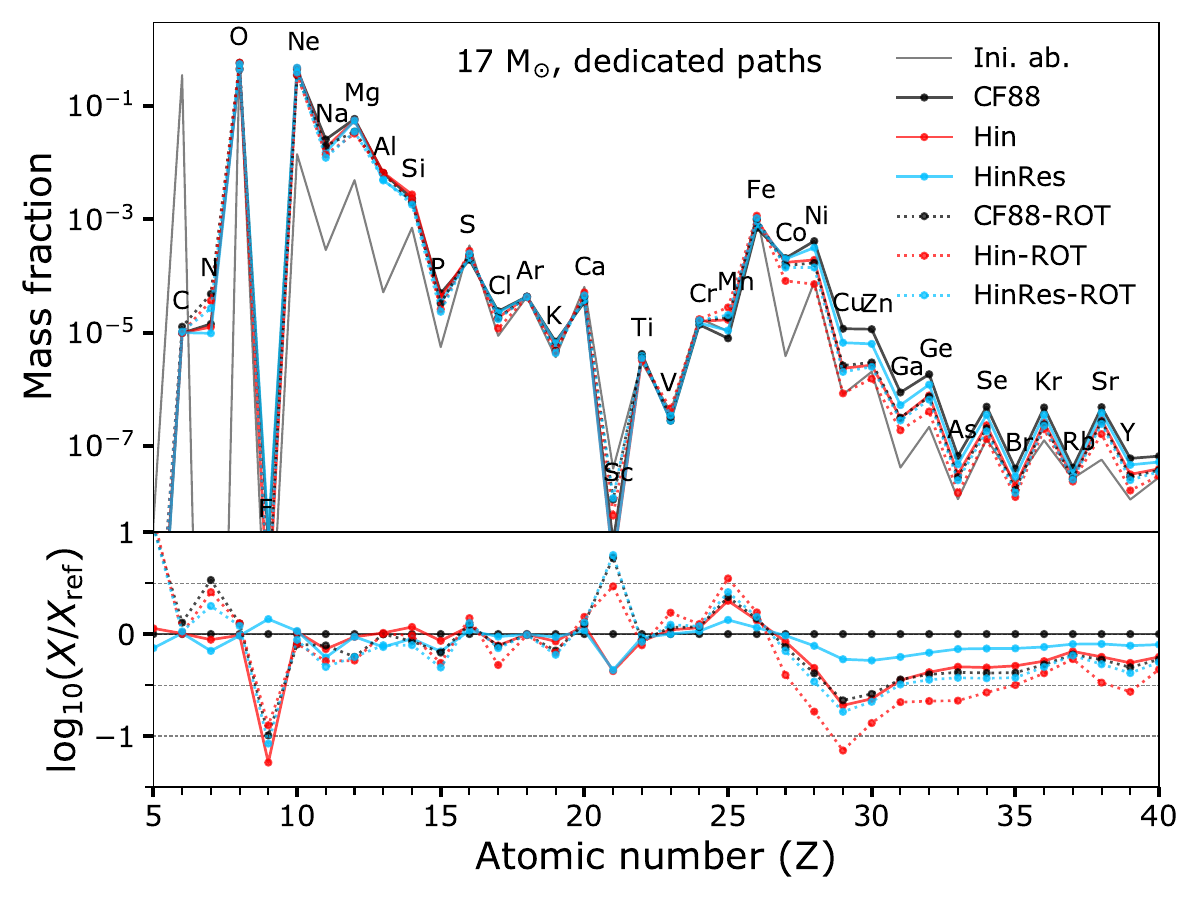}
         \includegraphics [width=90mm]{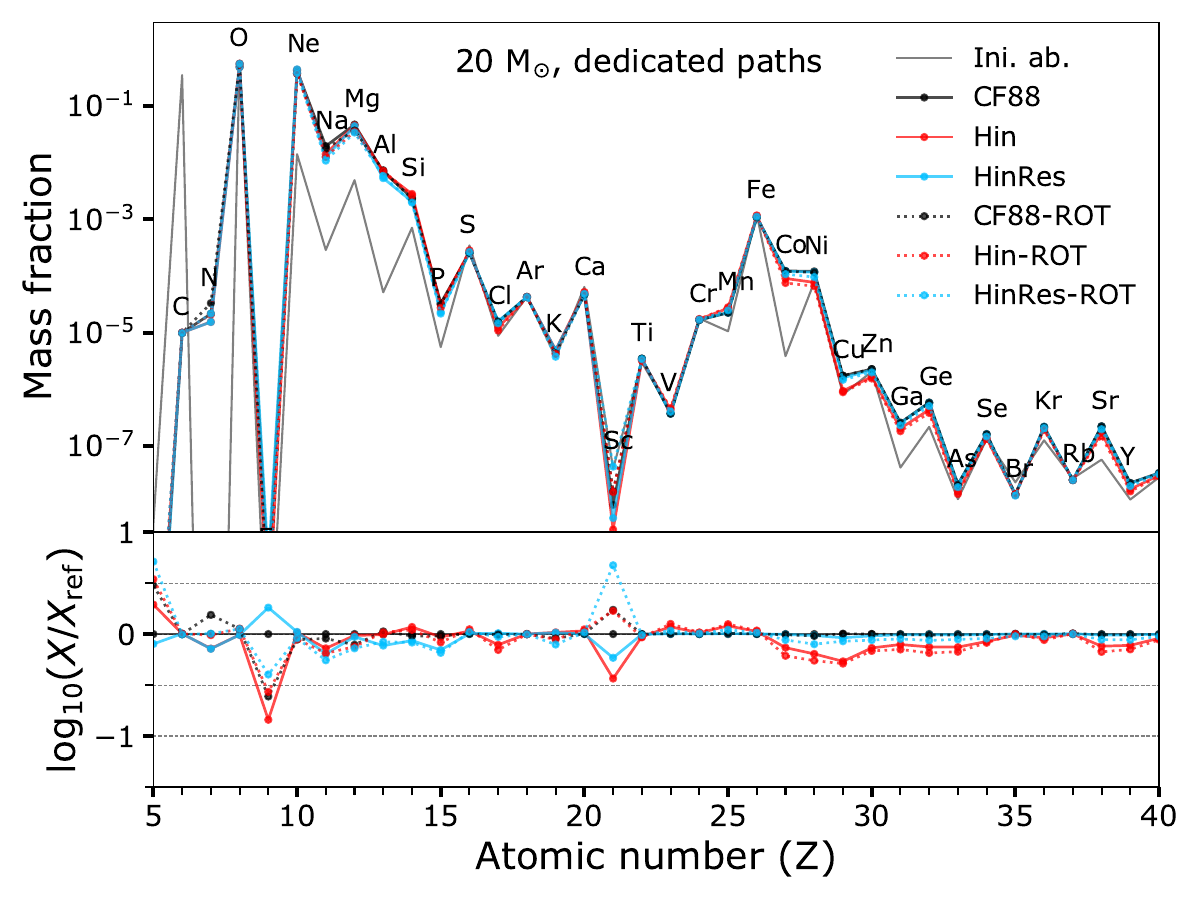}
         \caption{
         Same as Fig.~\ref{fig:nucleo_multi} but for the 17 (left panel) and 20 M$_{\odot}$ models (right panel), without (solid lines) and with rotation (dotted lines), and using dedicated paths.
         }
         \label{fig:nucleo_rot}
 \end{figure*}

In this part, we explore the effect of adding rotation that drives an internal mixing changing the temperature/density path and abundances all along the evolution. Upon C-ignition, abundances are consequently different in rotating models and can drive a different nucleosynthesis. As described above, the approximation of the one-zone model is less robust in that case as the rotation-induced mixing is not considered any more during the C-burning phase. Nevertheless, the C-burning burning timescale is smaller than the rotational mixing timescale as explained in Sect.~\ref{sect:nucl}. It means that rotation is not expected to alter significantly the structure of the star during this stage. The approximation then remains valid at first order. The main approximation remains that a one layer model does not capture all the complexity of multi zone stellar models. Figure~\ref{fig:nucleo_rot} shows the abundances at the end of the core C-burning phase for the 17 and 20 M$_{\odot}$ stars with (dotted lines) and without rotation (full lines) for the three nuclear references (colour-coded). \\ 

Including rotation in the 17~$M_{\odot}$ model weakens the s-process signature, no matter the nuclear rate reference (Fig.~\ref{fig:nucleo_rot}, left panel). 
This is again explained by the different C-burning lifetimes in these one-zone simulations. 
Overall, the central temperature during core C-burning is higher with rotation (Fig.~\ref{fig:crosstemp} and \ref{fig:crosstemp_rot}). It reaches $1.17$ and $1.24$~GK in the non-rotating and rotating 17~$M_{\odot}$ models, respectively. In our one-zone model, the carbon is therefore burnt quicker in the rotating case, leading to a shorter C-burning lifetime and therefore a smaller neutron exposure.
When including rotation in the 17~$M_{\odot}$ models, the neutron exposures are 1.7, 1.9 and 1.6 times smaller than when rotation is not included, for the CF88, HIN, and HINRES models, respectively.
This explains the smaller production of s-process elements for rotating models in Fig.~\ref{fig:nucleo_rot}. 

The differences are less important when considering the ($\rho,T$) paths of the 20~$M_{\odot}$ models (Fig.~\ref{fig:nucleo_rot}, right panel). The one-zone C-burning lifetimes of these six models range between about 145 and 500~yrs, and the neutron exposures between 0.06 and 0.09 mbarn$^{-1}$. It eventually results in a similar production of s-process elements.\\

\subsection{Discussion}
According to the results of the one-zone code, it is clear that the effect of rotation and of nuclear rates choice have a significant impact on the nucleosynthesis of the stars, especially for masses close to 17 M$_{\odot}$, with general underproduction of elements. While the change on the abundances is almost always weaker than 1 dex, it is nevertheless not negligible and exhibits an interesting mass dependence, with higher differences in the lowest mass stars of 15-17 M$_{\odot}$, where the rates are the most different, than in the more massive 20-25-30 M$_\odot$ (not shown). \\ For light elements, the main changes concern Na, slightly affected, while Mg, and Ne are not much affected. Also, as discussed in Sect.~\ref{sect:abond} about the dominant element in the core, we note that considering a more complete set of nuclear reactions, with notably the three exit channels of the $\rm ^{12}C + ^{12}\!C$ reaction, the abundance of $^{16}$O and $^{20}$Ne are similar in the core of each classical model, whereas the core is always $^{16}$O-dominated in rotating models. The abundance of $^{23}$Na remains at a mass fraction of about $10^{-2}$ for each model prediction\TDD{, close to the predicted values} in the full stellar GENEC models \TDD{which do not follow $^{23}$Na}.
\newline
In parallel, it is noteworthy mentioning that the s-process elements are affected by the nuclear reaction rates and by the rotation. 
Compared to the CF88 model, the lightest s-process elements are underproduced by about 0.5 and 0.2~dex in the HIN and HINRES models, respectively (Fig.~\ref{fig:nucleo_rot}, left panel). Including rotation reduces the production of light s-process elements by typically $0.2 - 0.5$~dex.
In Fig.~\ref{fig:nucleo_rot} (left panel), a peak is observed at $Z = 29$ which corresponds to an underproduction of light s-process elements by $\approx 0.5 - 1.0$ dex for the HIN and the HINRES rotating models compared to the CF88 non-rotating reference case.\\ 

The further exploration of s-process elements nucleosynthesis, especially during the next phases of stellar evolution (that will especially modify the core composition) and at different masses, is promising for a future study in that context. The study of nucleosynthesis in shells will the subject of a forthcoming study. In addition, we stress again that complete multi-zone stellar models, including complex processes like convection may affect the picture and should be tested in view of these results.

\section{Summary and outlook} 
\label{sect:CONCLUSION}
We computed models of massive non-rotating and rotating stars that include new nuclear reaction rates reference for the $\rm ^ {12}C+^{12}\!C$ fusion reaction. We explore two recent nuclear references according to the hypothesis of the Hindrance phenomenon (HIN model) and taking into account the presence of a resonance in the astrophysics region of interest (HINRES model), and we compare the results to the commonly used CF88 reference. We confirm that the change of nuclear reaction rates reference is impacting stars for the present studied mass-range (8-30 M$_{\odot}$), with an increase (decrease) of the temperature/density with the decrease (increase) of the cross section and nuclear reaction rates. In particular, HIN and HINRES models show shorter C-burning lifetimes compared to CF88 models. Interestingly, we highlight that stars are differently affected depending on their mass, especially in the case of the HINRES model that assumes a resonance localised in temperature. The resonance location results in more important changes at the beginning of the C-burning phase of the lowest mass stars and at the end of the C-burning phase for the most massive stars of our grid, as observed in Fig.~\ref{fig:rhot}, with a transition close to 20 M${_\odot}$ for classical models and close to 17 M${_\odot}$ for rotating models. \\ 
\newline
We explore two important mass range areas: 1) The mass range between 8 and 10 M$_{\odot}$, close to the M$_{\rm up}$ and M$_{\rm mas}$ critical mass limits, and 2) the maximum mass for which a convective core is formed. In the first case, we find that the new rates are not changing significantly the predicted value of M$_{\rm up} \approx 9-10$ M$_{\odot}$ and M$_{\rm mas} \approx 10-12$ M$_{\odot}$. However, for stars igniting carbon off-centre, the ignition occurs at a higher temperature and density in the HIN models than in the HINRES ones, themselves taking place at higher temperature and density than in the CF88 models. We cannot follow the carbon flash in current version of GENEC but we stress that the change of core and off-centre conditions, when changing the nuclear reaction rates, \TDD{might} impact the flame \TDD{characteristics as suggested by \citet{2023arXiv230300060D}}. The change of the rates is consequently significant for the definition of the critical mass limits. It is also a new constraint on the mass range driving the formation of hybrid CO-Ne core white dwarfs in place of ONe white dwarfs (between 8 and 10 M$_{\odot}$) that should be further explored. \\ In the second case, we demonstrate that the choice of the rates has an important impact on the threshold for the formation of a convective core during the C-burning phase. Comparing the results by \citet{2021ApJ...916...79C} using THM (Trojan horse method) rates with our results using HIN rates, we obtain a difference of more than 10 M$_{\odot}$ (from about 20 M$_{\odot}$ to about 32 M$_{\odot}$, respectively, in classical models), with the HINRES rates giving an intermediate value. We emphasize that the impact on the compactness and explosibility is then highly dependent on the choice of nuclear reaction rates and that this choice will impact the final fate of stars as well as the composition of the supernova ejecta (see for instance the relation between the CO core evolution and stellar remnants by \citet{2020MNRAS.499.2803P}). \\ We then advise the intermediate value of 24 M$_{\odot}$ predicted by our rotating HINRES models as the most reliable value. Indeed, it is based on the most realistic input physics, including rotation, and a nuclear model that shows the greatest similarities with the direct experimental results.\\
\newline
The stellar structural changes impact as well the chemical abundances in stars in both the HIN and the HINRES models compared to the original CF88 reference. It is mainly due to the change of convective core/shell size during evolution. Three initial mass regions can be identified for C-burning core, from rotating (classical) models: a first one at less than $\approx$ 22 M$_{\odot}$ ($\approx$ 20 M$_{\odot}$) where the CF88/HINRES models exhibits larger convective cores than the HIN models; a second region between $\approx$ 22 M$_{\odot}$ and $\approx$ 24 M$_{\odot}$ ($\approx$ 20 M$_{\odot}$ and $\approx$ 27 M$_{\odot}$) where the HIN models exhibit larger convective cores than the CF88/HINRES models; and finally a third region where the convective core is not formed for the CF88/HINRES models compared to the HIN models. A similar prediction is obtained for the carbon burning shells resulting in different chemical structures. \\ The combination of rotation-induced mixing and new rates is then impacting as well the nucleosynthesis and, especially, the production of s-process elements. The main production of s-elements is predicted during the \TDD{core} He-burning and shell C-burning phases, but a non-negligible contribution\TDD{, although small,} is also expected \TDD{from} the core C-burning phase. \TDD{Due to mixing, a part of the produced s-elements will contribute to the abundances outside the core} \citep[][]{2013ApJ...762...31P}. \TDD{The remaining s-elements in the core will then be destroyed during the next phases of evolution and the final explosion, contributing to the p-process \citep[the s-process elements will be the seeds of the p-process, e.g.][]{rayet95,rauscher02,arnould03,choplin22}.} At the end of core C-burning phase, an underproduction of s-process elements is predicted in the HIN models. The same effect is observed at a lower level in the HINRES models, compared to the CF88 models. It is mainly due to the lower burning lifetimes in these models decreasing the neutron exposure. Interestingly, we observe an important dependence to the mass for the level of the underproduction. Also, the rotational-induced mixing is accentuating the differences between the models. \\
\newline
New rate determinations of several important nuclear reactions have been done in recent years for the successive nuclear H-, He-, and C-burning fusion phases of stellar evolution, especially by the team of LUNA \citep{2018PrPNP..98...55B}. The update of the reaction rates could have an impact as it will impact the production/depletion of abundant elements like C and O, of light elements like Li, Be, and B \citep{Rapisarda2020}, but also the abundances of s-process elements that depend on the neutron density along evolution. The prediction of the $\rm ^ {12}C$ abundance at the end of the He-burning phase, that determines the C-burning phase, would, for instance, benefit of reducing the uncertainties over the fundamental reaction $\rm ^{12}C(\alpha,\gamma)^{16}O$ \citep[\textit{e.g.}][]{2001ApJ...558..903I,2013ApJ...762...31P,2017RvMP...89c5007D}. In addition, it has to be mentioned that several uncertainties were not considered in this work. On the nuclear side, the branching ratio between the channels is changing with energies/temperatures, which should increase the mass dependence. It was already shown by \citet{2013ApJ...762...31P} that changing the ratio between $\alpha$ and proton channel results in significant changes. On the stellar modelling side, we assume and explore only one rotation prescription and only one rotation velocity for all masses, but it is still today matter of debate on the prescription choice \citep[especially the shear coefficients, see for instance][]{2013LNP...865....3M,2021A&A...646A..48D}.\\
\newline
In a near future we plan to explore the consequences during the shell C-burning, known to be a main contributor of s-elements production \citep[\textit{e.g.}][]{1991ApJ...371..665R,1993ApJ...419..207R,2007ApJ...655.1058T}. In addition, the impact of multiple resonances for the $\rm ^{12}C+^{12}C$ reaction measured by the STELLA experiment (Nippert et al. 2024, in prep), as well as new data from recent and planed measurements on $\rm ^{12}C+^{16}O$ and $\rm ^{16}O+^{16}O$ nuclear reactions \citep{Yakovlev2006,Jiang2007,Jiang2011, Duarte2015, Fang2017} should be explored to improve the predictions on the final stages of evolution and fate of stars. \\ Finally, we stress that a better understanding on how the nuclear reaction rates are measured, as well as their uncertainties and potential improvements, should be taken into account when choosing nuclear reaction rates reference in stellar evolution codes.

\begin{acknowledgements}
This work was supported by the European Union (ChETEC-INFRA, project no. 101008324). The authors thank L. Siess and S. Martinet for helpful discussions, and the anonymous referee for constructive and helpful comments on the manuscript. T.D. thanks \TDD{K. Sieja for financial support for travel, J. Kerutt for helpful discussions, and} the Department of Astronomy of the University of Geneva for access to numerical resources. A.C. is a Postdoctoral Researcher of the Fonds de la Recherche Scientifique – FNRS. S.E. and G.M. acknowledge the support from the European Research Council (ERC) under the European Union’s Horizon 2020 research and innovation programme (grant agreement No 833925, project STAREX). This research has made use of NASA's Astrophysics Data System Bibliographic Services.
\end{acknowledgements}

%
%

\bibliographystyle{aa}
\bibliography{references}

\appendix
\section{\TDD{Nuclear network}}
\label{Annexe0}
\begin{table}[h]
    \centering
    \caption{\TDD{List of elements and isotopes considered in the models computed by the present version of GENEC stellar evolution code. Neutron are also considered.}}
    \begin{tabular}{c|c|c|c|c}
    \hline \hline
         Element & Element & Element & Element & - \\
         \hline 
         $^{1}$H & $^{16}$O & $^{25}$Mg & $^{52}$Fe & neutron\\
         $^{3}$He & $^{17}$O & $^{26}$Mg & $^{56}$Ni & \\
         $^{4}$He & $^{18}$O & $^{26}$Alg & & \\
         $^{7}$Be & $^{18}$F & $^{27}$Al & & \\
         $^{8}$B & $^{19}$F & $^{28}$Si & & \\
         $^{12}$C & $^{20}$Ne & $^{32}$S & & \\
         $^{13}$C & $^{21}$Ne & $^{36}$Ar & & \\
         $^{14}$C & $^{22}$Ne & $^{40}$Ca  & &  \\
         $^{14}$N & $^{23}$Na & $^{44}$Ti & & \\
         $^{15}$N & $^{24}$Mg & $^{48}$Cr  & & \\
         \hline
    \end{tabular}
    \label{tab:annexe0}
\end{table}
\section{Core evolution predictions}
\label{AnnexeA}
\begin{figure}[h]
    \centering
    \includegraphics [width=90mm]{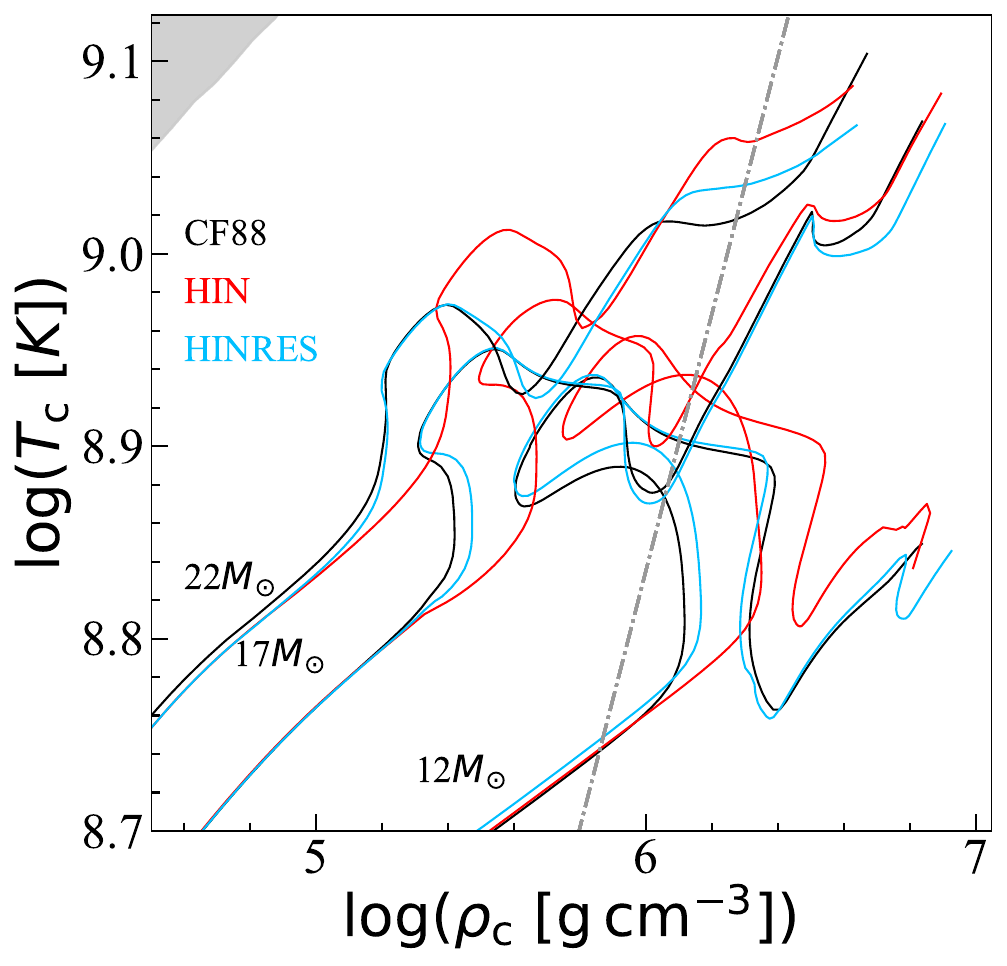}
    \caption{Evolution of T$_{\rm C}$ and $\rm \rho_{C}$ for classical models of the 12, 17, and 22 M$_{\odot}$ at solar metallicity, and for the three nuclear rates references (colour-coded). The grey dotted-dashed line indicates the limit between ideal gas (left) and degenerate gas (right). The grey shaded area indicates the pair instability domain $\rm e^+e^-$.}
    \label{fig:rhot_bis}
\end{figure}

\begin{figure}[h]
         \center
         \includegraphics [width=90mm]{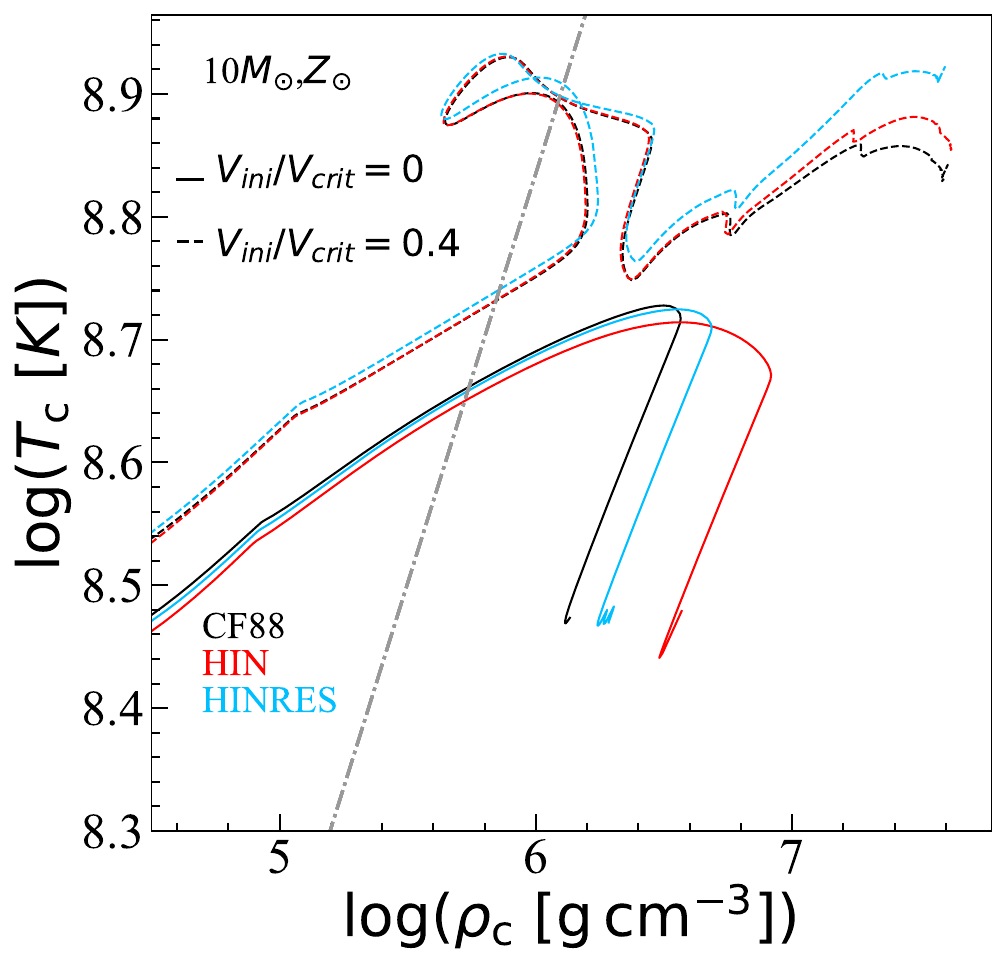}
         \caption{Evolution of T$_{\rm C}$ and $\rm \rho_{C}$ for classical (full line) and rotation (dashed line) models of the 10 M$_{\odot}$ at solar metallicity, and for the three nuclear rates references (colour-coded). The grey dashed-line indicates the limit between ideal gas (left) and degenerate gas (right).}
         \label{fig:Tcrhoc10}
 \end{figure}

Figure~\ref{fig:rhot_bis} shows the \TDD{central} temperature as a function of \TDD{central} density for additional models to Fig~\ref{fig:rhot} with the classical 12, 17 and 22 M$_{\odot}$, using the three different nuclear reaction rates described in the present paper: CF88, HIN, and HINRES.
\newline
 Figure~\ref{fig:Tcrhoc10} gives the \TDD{central} temperature as a function of \TDD{central} density for classical and rotating models of a 10 M$_{\odot}$, using different nuclear reactions rates references. Classical models predict an off-centre C-ignition while rotating models, hotter, predict a non-degenerate core C-ignition.
 
\end{document}